\documentclass[a4paper,fleqn,usenatbib]{mnras}

\usepackage{mathptmx}
\usepackage{txfonts}

\usepackage[T1]{fontenc}
\usepackage{ae,aecompl}

\usepackage{graphicx}	

\usepackage{amsmath}	
\usepackage{amssymb}	

\usepackage{pdflscape}
\usepackage[utf8]{inputenc}
\usepackage{xfrac}
\usepackage{color}

\title[Do A--type stars flare?]{Do A--type stars flare?}

\author[M. G. Pedersen et al.]{M. G. Pedersen,$^{1,2}$\thanks{E-mail: maygade.pedersen@kuleuven.be}
V. Antoci,$^{1}$
H. Korhonen,$^{3}$ 
T. R. White,$^{1}$
J. Jessen-Hansen,$^{1}$
\newauthor 
J. Lehtinen,$^{4}$
S. Nikbakhsh,$^{5}$
and J. Viuho$^{3,6}$
\\
$^{1}$Stellar Astrophysics Centre, Department of Physics and Astronomy, Aarhus University, 120 Ny Munkegade, Building 1520,\\ 8000 Aarhus C, Denmark\\
$^{2}$Institute of Astronomy, Department of Physics and Astronomy, KU Leuven, Celestijnenlaan 200D, 3001 Leuven, Belgium\\
$^{3}$Dark Cosmology Centre, Niels Bohr Institute, University of Copenhagen, Juliane Maries Vej 30, 2100 Copenhagen \O, Denmark\\
$^{4}$Max-Planck-Institut für Sonnensystemforschung, Justus-von-Liebig-Weg 3, 37077 Göttingen, Germany\\
$^{5}$Department of Radio Science and Engineering, Aalto University, Otakaari 5A, Espoo, Finland\\
$^{6}$Department of Physics, PO Box 64, 00014 University of Helsinki, Finland\\
}

\date{Accepted 2016 December 8. Received 2016 December 8; in original form 2016 September 2}

\pubyear{2016}

\begin{document}
\label{firstpage}
\pagerange{\pageref{firstpage}--\pageref{lastpage}}
\maketitle

\begin{abstract}
For flares to be generated, stars have to have a sufficiently deep outer convection zone (F5 and later), strong large--scale magnetic fields (Ap/Bp--type stars) or strong, radiatively driven winds (B5 and earlier). Normal A--type stars possess none of these and therefore should not flare. Nevertheless, flares have previously been detected in the \emph{Kepler} lightcurves of 33 A--type stars and interpreted to be intrinsic to the stars. Here we present new and detailed analyses of these 33 stars, imposing very strict criteria for the flare detection. We confirm the presence of flare-like features in 27 of the 33 A--type stars. A study of the pixel data and the surrounding field-of-view (FOV) reveals that 14 of these 27 flaring objects have overlapping neighbouring stars and 5 stars show clear contamination in the pixel data. We have obtained high-resolution spectra for 2/3 of the entire sample and confirm that our targets are indeed A-type stars. Detailed analyses revealed that 11 out of 19 stars with multiple epochs of observations are spectroscopic binaries. Furthermore, and contrary to previous studies, we find that the flares can originate from a cooler, unresolved companion. We note the presence of H$\alpha$ emission in eight stars. Whether this emission is circumstellar or magnetic in origin is unknown. In summary, we find possible alternative explanations for the observed flares for at least 19 of the 33 A--type stars, but find no truly convincing target to support the hypothesis of flaring A--type stars. 
\end{abstract}



\section{Introduction}

In the year 1859, R.C. Carrington and R. Hodgson detected the first and largest known solar flare \citep{Carrington1859,Hodgson1859}, which later came to be known as the \emph{Carrington Event}. Such highly energetic events occur in the solar atmosphere when magnetic field lines reconnect to a lower energy configuration \citep[e.g.][]{Charbonneau2010}. In this process a vast amount of magnetic energy is released in the form of accelerated particles and electromagnetic radiation, covering the entire wavelength spectrum \citep[e.g.][]{Shibata2011}. Solar flares typically have energies between $10^{28} - 10^{32}$ erg and durations of approximately $15 \ \text{min}$ to $2.5 \ \text{h}$ \citep[e.g.][]{Shibata2011}. In comparison, flares on active M-- and K--dwarfs are known to have flare energies 10-1000 times larger than solar flares \citep[e.g.][]{Walkowicz2011}. 

Flares in late--type stars (F-M) are believed to be generated through the same underlying mechanism as solar flares. In such stars, the magnetic fields are formed by a dynamo, which requires the presence of a convective envelope in order to operate \citep[e.g.][]{Charbonneau2010}. This convective envelope has to be sufficiently deep for the magnetic fields to get strong enough \citep[$\geq 10 \ \text{kG}$ for the Sun, e.g.][and references therein]{Charbonneau2010} to rise and emerge into the stellar atmosphere, and thereby lead to the formation of flares. Observational evidence combined with theoretical predictions of the onset of deep outer convection zones imply that dynamo-generated magnetic fields first occur at spectral type $\sim$F5 \citep[e.g.][and references therein]{Rosner1980}. 

Early--type stars have only shallow convective envelopes and are therefore incapable of producing flares through known dynamo mechanisms. Some of these stars have, however, coincidentally been found to show X-ray flare activity. Examples of such stars are $\lambda$ Eridani \citep[B2e,][]{Smith1993}, MWC 297 \citep[Herbig Be,][]{Hamaguchi2000}, $\sigma$ Ori E \citep[B2p,][]{Groote2004,Sanz-Forcada2004}, IQ Aur \citep[A0p,][]{Robrade2011} and the massive binary system $\eta$ Carinae \citep{Moffat2009}. In the case of MWC 297, the X-ray flaring has since been found to originate from a late--type star in the same field-of-view \citep{Vink2005}. For the other stars, the observed events can be explained through other mechanisms, as summarized below.

Ap-- and Bp--type stars are characterised by having chemical peculiarities which arise due to the presence of strong and stable, large-scale magnetic fields of the order of $200 \ \text{G} - 30 \ \text{kG}$ \citep[e.g.][]{Braithwaite2014,Power2007}. One explanation for the origin of these fields is the \emph{fossil field theory}, according to which the fields are remnants of the primordial fields of the interstellar gas from which the stars are formed \citep[e.g.][and references therein]{Braithwaite2014,Dudorov2015}. \citet{Townsend2005} proposed that strong, large-scale magnetic fields can confine and guide the stellar winds along the field lines towards the magnetic equator, where material starts to accumulate. When the centrifugal force acting on this trapped material becomes sufficiently large, the matter is ejected outwards, dragging along magnetic field lines until they reconnect and snap back towards the star. This is known as a \emph{magnetic breakout event}. \citet{Owocki2007} suggested that this mechanism can explain the X-ray flares observed in $\sigma$ Ori E and $\theta^2$ Ori A (O9.5Vpe). 

Another mechanism is related to stellar winds which become radiatively driven and hence increasingly stronger towards early--type stars. If two massive stars are placed in a binary system, the colliding winds of the stars will lead to X-ray emission varying with the orbital phase \citep[e.g.][]{Capelli2011}. Winds in hot stars are composed of many smaller and fewer larger, turbulent clumps. One explanation for the observed flaring in $\eta$ Carinae is that they occur when the larger clumps enter the colliding wind zone between the stars, while the short term emission is caused by the more frequent smaller clumps \citep{Moffat2009}. The $\eta$ Carina system consists of a Luminous Blue Variable (LBV) primary star ($M \sim 90 \ M_\odot$) and a hot, fast wind secondary companion ($M \sim 30 \ M_\odot$) which is believed to be either an evolved O--type or a Wolf-Rayet star \citep[e.g.][]{Moffat2009,Verner2005,Pittard2002}. LBVs have a large number of clumps in their winds, which makes this scenario a possible explanation for the observed flares in this system. The transition between weak and strong winds takes place on the main-sequence for temperatures between $T_\text{eff} \approx 14000-18000 \ \text{K}$, corresponding to stars with spectral types between B3 and B5 \citep[e.g.][]{Stelzer2005}.

Finally, stars of spectral type F5 to B5 are not expected to flare as they occupy the activity quiet spectral range. A--type stars are situated right in the middle of this range. These stars have only weak stellar winds and shallow convective envelopes of the order of $1-3\%$ of the stellar radius. In comparison, the outer convection zone of the Sun has a thickness of $0.3R_\odot$. Normal A--type stars only have weak, global magnetic fields as detected for Vega \citep[$0.6 \pm 0.3 \ \text{G}$, ][]{Lignieres2009}, Sirius \citep[$0.2 \pm 0.1 \ \text{G}$, ][]{Petit2011} and the $\delta$ Sct star HD 188774 \citep{Neiner2015}. Therefore, none of the above mentioned flare mechanisms should be operating in these stars. 

In spite of this, \citet{Balona2012,Balona2013} detected flare-like events in the \emph{Kepler} lightcurves of 33 A--type stars through a visual inspection of the lightcurves. Possible and likely explanations for such observed flares are: 1) the spectral classification of the stars is wrong and the stars are in fact of a spectral type where known flare mechanisms may operate, 2) they originate from or are caused by an interaction between the A--type star and a cool, unresolved companion, 3) caused by infalling bodies from a possible circumstellar disk, 4) purely instrumental effects, or 5) due to contamination from a nearby star. \citet{Balona2012,Balona2013} find all these explanations unlikely, although not all were tested in detail, and attribute the flares to the A stars, contradicting theory.

Here we carry out detailed analyses of each of the supposedly 33 flaring A--type stars listed by \citet{Balona2012,Balona2013}. In order to conclude that the events originate solely from the A--type stars, all the above mentioned possible and more likely explanations have to be ruled out first. In Section \ref{Sec:Photometry} we analyse the \emph{Kepler} data by carrying out a stringent flare detection in an attempt to reproduce the results of \citep{Balona2012,Balona2013}. For each of the detected flares we determine basic flare parameters. We also consider the long-term stellar variability in order to determine if the stars show signatures of spots or binarity. In Section \ref{Sec:PixelData} we consider the pixel data of the stars found to be flaring in order to investigate if the lightcurves are contaminated. In Section \ref{Sec:Spectroscopy}, we assign each star a spectral type, determine radial velocity shifts and look for evidence of circumstellar material. We discuss the results in Section \ref{Sec:Discussion} and give the final conclusions in Section \ref{Sec:Conclusions}.

\section{Photometric analysis}\label{Sec:Photometry}

The initial step of the analysis is carried out by considering the \emph{Kepler} \citep[e.g.][]{Koch2010} data of all 33 flaring A--type stars listed by \citet{Balona2012,Balona2013}. \emph{Kepler} lightcurves are taken at two different cadences. The long cadence (LC) and the short cadence (SC) data have a sampling of $30 \ \text{min}$ and $1 \ \text{min}$ respectively \citep{ChristiansenVanCleve2011}. Because most flares have durations of the order of minutes, the best way to study flares and their characteristics would be to look at the SC data. However, few of the 33 \textit{Kepler} targets were observed in SC and usually for not more than 1 month. In contrast, most observed stars will have several years of LC data available, with a maximum of 4.5 yr, corresponding to the \emph{Kepler} mission life time. Stars with SC data will also have LC data covering the same period.

The \emph{Kepler} data were acquired in segments of three months, referred to as quarters, after the spacecraft performed its quarterly roll \citep{VanCleve2009}. The quarters are numbered from Q0 to Q17 accordingly, where Q0 corresponds to the commissioning phase of $10 \ \text{d}$ and Q1 only has a length of one month. In this analysis, we now consider the entire \emph{Kepler} data set from Q0-Q17 in comparison to \citet{Balona2012,Balona2013}, who only used the available Q0-Q6 and Q0-Q12, respectively. 

The data were downloaded via the KASOC homepage\footnote{http://kasoc.phys.au.dk}. Because the \emph{Kepler} pipeline is optimised for planet detections and might remove genuine flare signals, we chose to use the raw Simple Aperture Photometry (SAP) data. The lightcurves were corrected for instrumental trends by removing the time intervals affected by loss of fine-pointing, quarterly rolls and monthly downlinking of the data. Jumps in the measured flux are adjusted to the same level by subtracting by the flux difference before and after the jump. Finally, the long term instrumental trends are corrected for by dividing by a 1st-5th order polynomial fit to the dataset. The exact polynomial order is determined individually from star to star and from quarter to quarter. With this correction we end up with the relative flux, $F_\text{rel}$, representing the stellar variability as a function of time.

\subsection{Flare detection}\label{Sec:FlareDetection}

In photometry, flares are seen as a sudden increase in the brightness of the star followed by a slow exponential decay which may last between a couple of minutes to a few hours. Therefore, when analysing the lightcurves we are interested in detecting brightening events which differ significantly from the normal stellar variability. Since most stars only have LC data available and the number of flares listed by \citet{Balona2012,Balona2013} are from the LC data only, we will base our search on these lightcurves as well. Based on the characteristics mentioned above and the data type we are using, we define three criteria for the flare detection process:

\begin{enumerate}
	\item The flare must consist of a sudden spike in flux with at least three data points in LC showing a slow exponential decay above a 3$\sigma_\text{std}$ detection limit.
	\item The duration must be less than $10 \ \text{h}$.
	\item If SC data covering the range of the flare are available, the LC flare must be fully resolved in SC.
\end{enumerate}

Similar criteria were used by \citet{Walkowicz2011} to detect flares in the Q1 LC data of M and K dwarfs. 

The first detection criterion is used to rule out instrumental effects due to cosmic rays and outliers. While the cadence of the LC data makes it possible that single point outliers could in fact be short-lived flares, it is impossible to distinguish between this scenario and simple outliers. To carry out a robust flare detection, we require that the flare candidates consist of at least three data points in LC. As a consequence the shortest detectable flare has a duration of $\sim 1.5  \ \text{h}$. This is not a problem because flares usually have durations of a few minutes to several hours. In the case of M-- and K--dwarfs, for example, \citet{Walkowicz2011} found the median flare duration to be $2-6 \ \text{h}$. Therefore we set the second flare detection to require that the duration of a possible flare event has to be less than $10 \ \text{h}$. Furthermore, flares detected in LC should also be observable in the SC data. Hence the third and final detection criterion.


\begin{figure*}
	\includegraphics[width=0.9\linewidth]{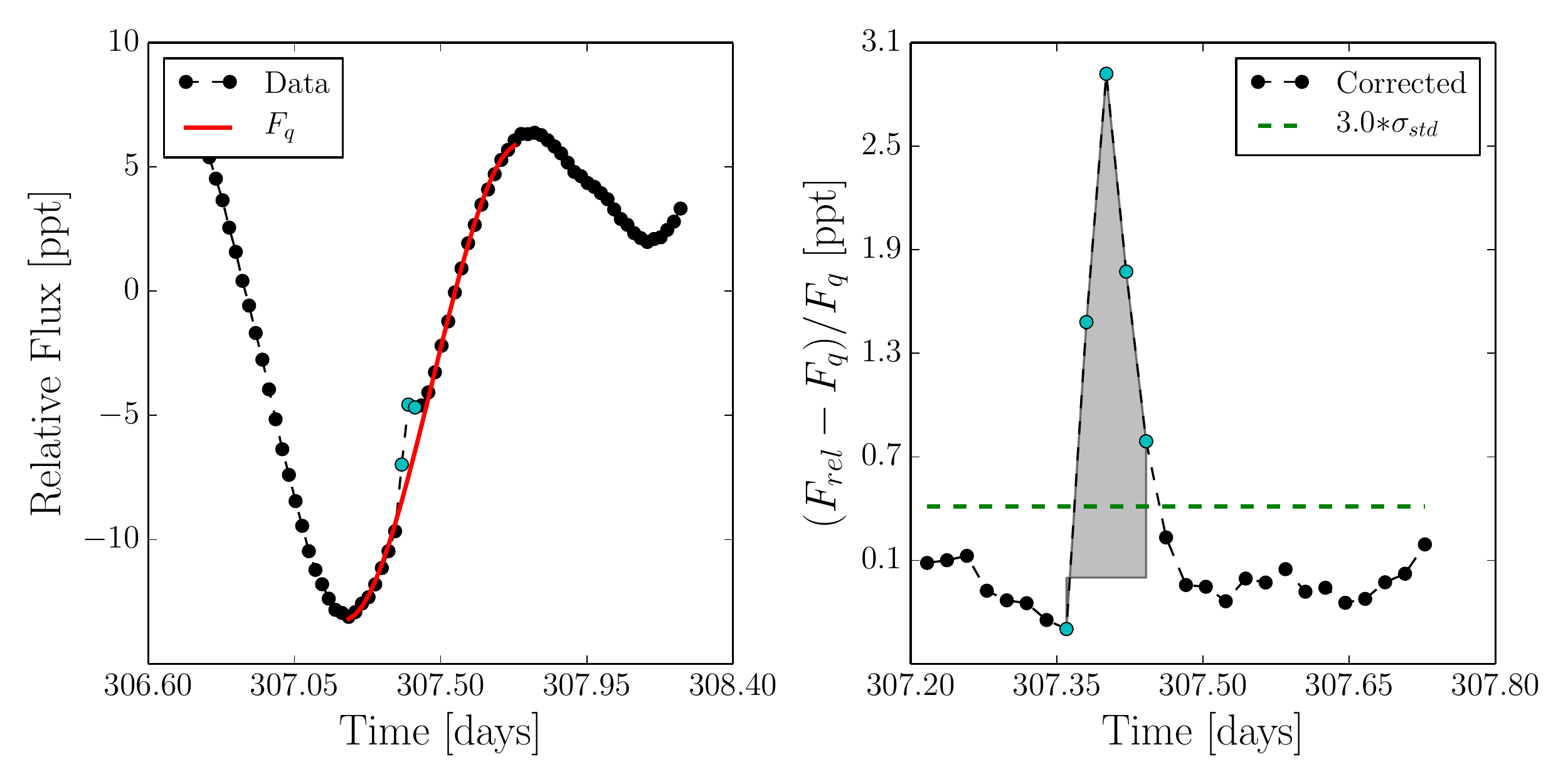}
    \caption{Illustration of the flare detection procedure for a flare found in the lightcurve of KIC 7097723. The relative flux is plotted in \emph{parts-per-thousand} (ppt). \emph{Left panel}: Suspected flare data points are marked by cyan filled circles and excluded from the polynomial fit (red curve) of the quiescent flux $F_\text{q}$ in an interval around the flare. \emph{Right panel}: Resulting lightcurve after quiescent flux correction. Green dashed line shows the $3\sigma_\text{std}$ detection limit. Data points above this detection limit are marked as flare data points (cyan filled circles) as well as the data point just prior to the flare. The grey shaded region is the area beneath the flare which when integrated corresponds to the \emph{photometric equivalent width} $EW_\text{phot}$, see text.}
    \label{fig:FlareDetection1}
\end{figure*}

\begin{figure*}
	\includegraphics[width=0.9\linewidth]{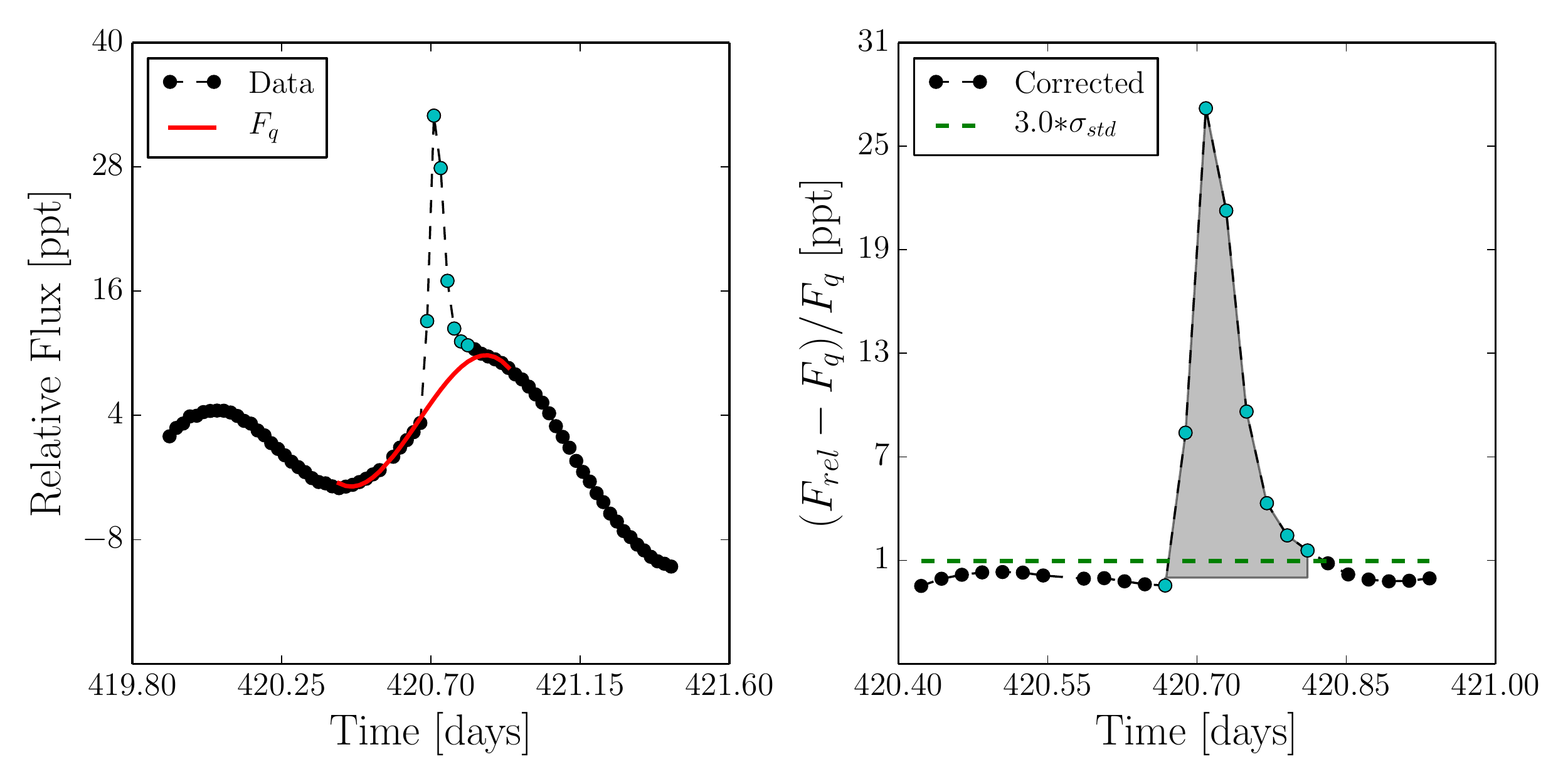}
    \caption{Same as Fig. \ref{fig:FlareDetection1} but for a more prominent flare found in the lightcurve of KIC 7097723.}
    \label{fig:FlareDetection}
\end{figure*}

Possible flare candidates are identified by visual inspection of the lightcurves and subjected to the three detection criteria mentioned above after correcting for the quiescent flux, $F_\text{q}$. This is the flux the star would be expected to emit in the absence of the flare. The process is illustrated in Fig. \ref{fig:FlareDetection1} and \ref{fig:FlareDetection}. The quiescent flux correction is done by fitting a fourth order polynomial (red curve) in an interval around the flare after excluding the suspected flare data points (blue filled circles) as illustrated in the left panel of Fig. \ref{fig:FlareDetection1} and \ref{fig:FlareDetection}. The remaining scatter after the quiescent flux correction is due to noise and indicates how well the fourth order polynomial fits the stellar variability, as well as the intrinsic stellar and instrumental noise. The resulting corrected lightcurve is shown in the right panel of Fig. \ref{fig:FlareDetection1} and \ref{fig:FlareDetection}.

For the first detection criterion we calculate the standard deviation, $\sigma_\text{std}$, of the scatter, excluding the cyan data points in the left panels in Fig. \ref{fig:FlareDetection1} and \ref{fig:FlareDetection}. For a brightening to be marked as a flare, at least three consecutive data points have to show an exponential decay above a $3\sigma_\text{std}$ detection limit (green dashed line in the right panel of Fig. \ref{fig:FlareDetection1} and \ref{fig:FlareDetection}). The cyan filled circles in the right panels mark the data points fulfilling this criterion, including the data point just prior to the flare. This additional data point is used for the flare parameter determination in the next section.

\begin{table*}
	\caption{Results from flare detection and Fourier analysis. The first column gives the \emph{KIC} number of the star, \emph{Kp} is the \emph{Kepler} magnitude, T$_{\text{eff}}$ is the effective temperature obtained from the \emph{Kepler Input Catalogue} and has an uncertainty of at least $\pm 200 \ \text{K}$ \citep{Brown2011}, \emph{LC} and \emph{SC} are the number of quarters with LC and SC data available, $N_\text{Balona}$ is the number of flares detected by \citet{Balona2012} (superscript $i$) and \citet{Balona2013} (superscript $ii$), $N_\text{Flares}$ is the number of flares detected in this work and $N_\text{Spikes}$ is the number of spikes, log$EW_\text{phot}$ is the calculated \emph{photometric equivalent width} of the most energetic flare and log $\Delta F/F$ is the corresponding amplitude and $t_{wait}$ is the average waiting time between flares.}
	\label{Tab:Flares}
	\begin{tabular}{cccccccccccccccc}
	\hline
	\text{KIC}		& \text{Kp} 	&	\text{T$_{\text{eff}}$}	& \text{LC} 	&	\text{SC}		& \text{$N_\text{Balona}$}	&	\text{$N_\text{Flares}$}	&	\text{$N_\text{Spikes}$}	&	\text{log$EW_\text{phot}$}	& \text{log $\Delta F/F$}	&	\text{$t_{wait}$}	\\
	\text{}	& \text{[mag]}	&	\text{[K]}	&	\text{}  &	\text{}	& \text{}	&	\text{}	&	\text{}	&	\text{[h]}	&	&	\text{[d]}\\
	\hline
	3974751	&	11.4	&	9300	&	15	&	0	&	60$^i$	&	0	&	3	&	-	&	-	&	-	\\
  4472809	&	14.0	&	8400	&	14	&	0	&	3$^i$	&	19	&	2	&	-1.742	&	-2.003	&	64.9	\\
  4773133	&	12.1	&	9400	&	18	&	0	&	2$^i$	&	2	&	0	&	-2.685	&	-2.952	&	1358.6\\
  5113797	&	9.1	&	8100	&	15	&	1	&	24$^i$	&	3	&	4	&	-2.820	&	-2.812	&	458.1\\
  5201872	&	9.5	&	7900	&	15	&	1	&	4$^i$	&	6	&	2	&	-2.459	&	-2.553	&	263.3\\
  5213466	&	13.1	&	8300	&	18	&	0	&	14$^i$	&	5	&	2	&	-2.282	&	-2.468	&	312.3\\
  5559516	&	8.7	&	9300	&	18	&	0	&	5$^i$	&	2	&	0	&	-3.973	&	-3.930	&	755.2\\
  5870686	&	9.9	&	7500	&	18	&	1	&	26$^i$	&	6	&	4	&	-2.630	&	-2.899	&	257.9\\
  5898780	&	12.8	&	8500	&	17	&	0	&	8$^i$	&	1	&	0	&	-2.283	&	-2.289	&	-\\
  6219684	&	12.2	&	9400	&	18	&	0	&	6$^{ii}$	&	8	&	13	&	-1.488	&	-1.654	&	188.5\\
  6451234	&	12.8	&	8100	&	18	&	0	&	4$^i$	&	2	&	0	&	-2.922	&	-2.811	&	1170.9\\
  7047141	&	12.7	&	8200	&	18	&	0	&	8$^{ii}$	&	9	&	4	&	-1.171	&	-1.436	&	118.8\\
  7097723	&	11.7&	8400	&	14	&	0	&	11$^i$	&	25	&	8	&	-1.397	&	-1.557	&	51.8\\
  7809801	&	14.6	&	8100	&	2	&	0	&	9$^{ii}$	&	2	&	0	&	-2.033	&	-1.942	&	13.3	\\
  7974841	&	8.2	&	8900	&	18	&	1	&	22$^{ii}$	&	0	&	1	&	-	&	-	&	-	\\
  7978512	&	12.3	&	9300	&	18	&	0	&	21$^i$	&	8	&	4	&	-2.022	&	-2.293	&	145.0\\
  8044889	&	12.6	&	7700	&	18	&	0	&	1$^{ii}$	&	18	&	8	&	-1.839	&	-2.095	&	82.0	\\
  8351193	&	7.6	&	8500	&	17	&	1	&	14$^i$	&	1	&	0	&	-3.627	&	-3.675	&	-	\\
   8367661	&	8.7	&	8500	&	18	&	0	&	23$^{ii}$	&	0	&	2	&	-	&	-	&	-\\
  8686824	&	10.7	&	9000	&	18	&	0	&	30$^i$	&	1	&	0	&	-2.464	&	-2.744	&	-\\
  8703413	&	8.7	&	7700	&	18	&	1	&	8$^i$	&	0	&	0	&	-	&	-	&	-	\\
  9216367	&	12.1	&	7800	&	18	&	1	&	31$^i$	&	13	&	3	&	-1.850	&	-2.113	&	113.1\\
  9782810	&	9.5	&	9300	&	15	&	0	&	3$^{ii}$	&	1	&	0	&	-3.409	&	-3.397	&	-\\
  10082844	&	13.7	&	9400	&	17	&	0	&	7$^i$	&	7	&	2	&	-1.735	&	-1.840	&	146.5\\
  10489286	&	11.8	&	9400	&	15	&	0	&	63$^i$	&	19	&	11	&	-2.454	&	-2.545	&	77.7	\\
  10817620	&	14.0	&	8000	&	14	&	0	&	2$^{ii}$	&	8	&	2	&	-2.081	&	-2.168	&	186.4\\
  10971633	&	11.5	&	9000	&	18	&	0	&	7$^i$	&	5	&	4	&	-2.585	&	-2.630	&	219.7\\
  10974032	&	8.4	&	9000	&	18	&	1	&	19$^{ii}$	&	3	&	2	&	-3.784	&	-3.665	&	462.1\\
  11189959	&	8.2	&	9300	&	18	&	1	&	38$^i$	&	17	&	3	&	-2.509	&	-2.537	&	61.2	\\
  11236035	&	14.6	&	7900	&	4	&	0	&	4$^i$	&	1	&	2	&	-1.020	&	-1.229	&	-	\\
  11443271	&	7.5	&	8200	&	14	&	1	&	6$^{ii}$	&	0	&	0	&	-	&	-	&	-	\\
  11600717	& 	7.6	&	7900	&	1	&	1	&	3$^{ii}$	&	0	&	0	&	-	&	-	&	-	\\
  12061741	&	8.6	&	9200	&	17	&	0	&	1$^{ii}$	&	1	&	0	&	-3.519	&	-3.558	&	-	\\
	\hline
	\end{tabular}
\end{table*}

We compare the timings of all detected flares in this paper and discard the detections if the events overlap with a) another flare, b) excessive scattering, c) one or more outliers or d) small jumps, in at least one other star. Simultaneous flare events in several stars can occur but it is also very likely that these may be caused by instrumental effects. Therefore we prefer to exclude such flares from our analyses. As an example, KIC 8351193 is a star that shows a lot of scattered data points and outliers from both above and below its lightcurve. Therefore, we find it more likely that the single detected flare event in this star is of instrumental origin rather than a genuine stellar flare.

Results from the flare detection are listed in Table \ref{Tab:Flares}. $N_\text{Balona}$ gives the number of flares detected by \citet{Balona2012,Balona2013} and $N_\text{Flares}$ is our corresponding result. By comparing these values it is immediately evident that there is a large discrepancy between the numbers. For 7 stars we detect more flares than \citet{Balona2012,Balona2013}, which given the amount of data we are using is to be expected. For 23 stars we detect fewer flares, out of which 6 stars are not found to be flaring at all. This is surprising as we include more quarters in our analysis, corresponding to additional $2 \sfrac{3}{4}$ yr and $1\sfrac{1}{2}$ yr worth of data compared to \citet{Balona2012,Balona2013} respectively and should therefore be detecting more flares. $N_\text{Spikes}$ lists the number of events where we detect three data points above the $3\sigma_\text{std}$ limit, but they fail to show an exponential decay. Such features might still be flares, but even including these we cannot account for the discrepant values. 

While \citet{Balona2012,Balona2013} also identify the flares through visual inspection, they do not use any specific flare detection criteria. As explained in \citet{Balona2012,Balona2013}, the detection includes single point outliers through the argument that since most flares have durations less than half an hour, single point outliers could in fact be flares. Further \citet{Balona2012,Balona2013} argues that after taking the overall noise level of the lightcurves into consideration, such outliers may be included as potential flare candidates. While this is indeed a valid argument, there is no way to distinguish between this scenario and instrumental outliers without the addition of SC data. Not identifying single outliers as flares is common practice even when analyzing stars which are known to be intrinsically flaring \citep[e.g.][]{Walkowicz2011}.

\subsection{Flare parameters}\label{Sec:FlareParameters}

Following the procedure in \citet{Walkowicz2011} we calculate the \emph{photometric equivalent width}, $EW_\text{phot}$, as a measurement for the flare energy,

\begin{equation}
EW_\text{phot} = \int \frac{F_\text{rel}-F_\text{q}}{F_\text{q}} dt,
\end{equation}

\noindent where $F_\text{rel}$ is the relative flux of the star and $F_\text{q}$ is the fitted quiescent flux. $EW_\text{phot}$ gives the time interval over which the star emits as much energy as during the flare and is hence given in units of time, and is the integrated flux of the flare represented by the grey shaded area in the right panel of Fig. \ref{fig:FlareDetection1} and \ref{fig:FlareDetection}. The first data point just prior to the flare is included in this calculation as well as in determining the duration. This was done in order to account for the unknown starting time of the flare. 

We also determine the flare amplitude as

\begin{equation}
\text{Amplitude} = \frac{\Delta F}{F} = \left(\frac{F_\text{rel}-F_\text{q}}{F_\text{q}}\right)_\text{peak},
\end{equation}

\noindent which is the peak flux count in the right panel of Fig. \ref{fig:FlareDetection1} and \ref{fig:FlareDetection}. The final determined parameter is the waiting time between the flares, $t_\text{wait}$. $EW_\text{phot}$, amplitudes of the largest flares as well as the average waiting times between flares are listed in Table \ref{Tab:Flares}.

\begin{figure}
	\includegraphics[width=\linewidth]{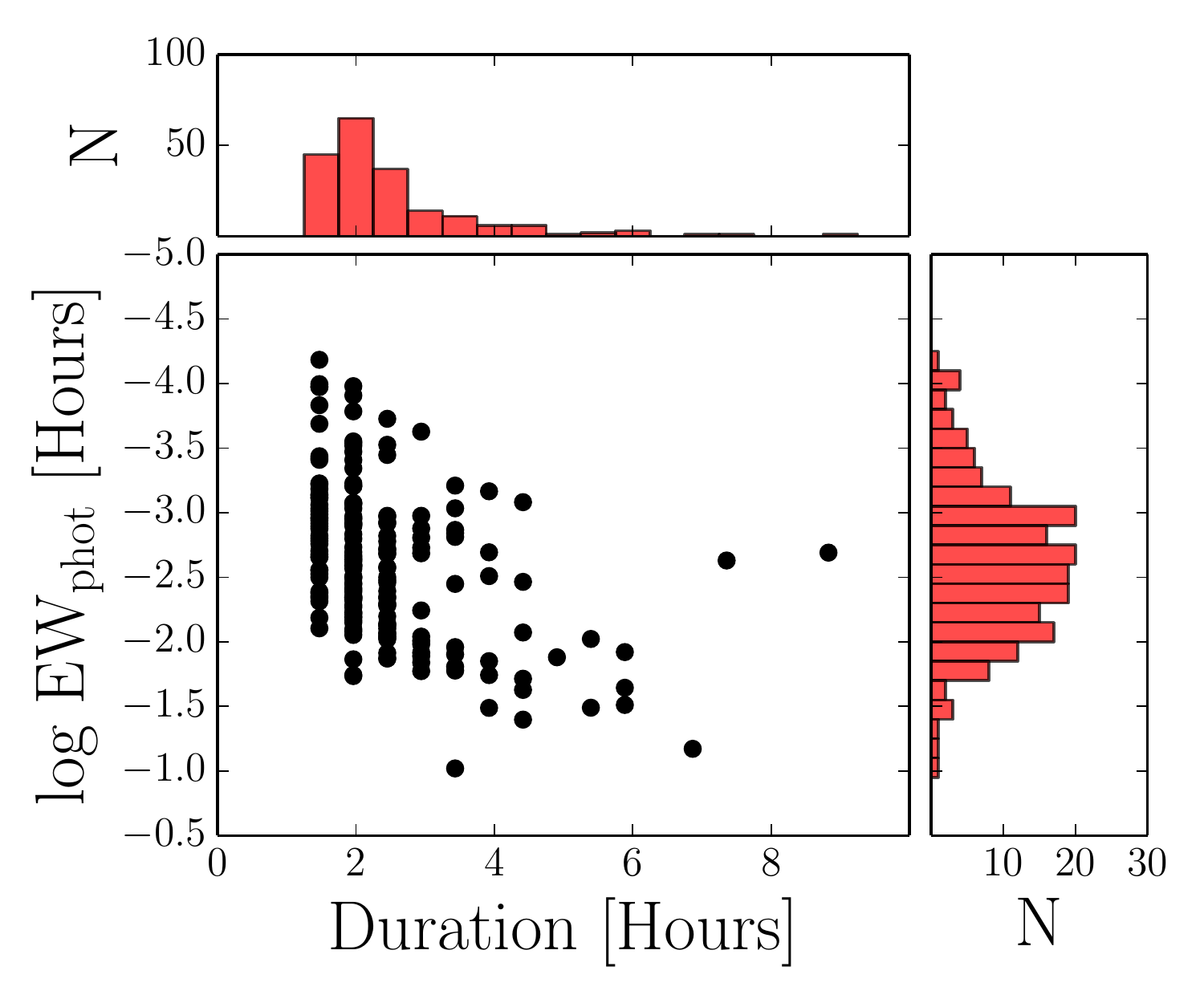}
    \caption{Flare energy ($EW_\text{phot}$) as a function of duration for all detected flares for all 27 A--type stars found to be flaring in this work. \emph{Upper panel}: histogram of the number of flares having a given duration. \emph{Right panel}: histogram of the number of flares having a given energy.}
    \label{fig:EWvsDuration}
\end{figure}

\begin{figure}
	\includegraphics[width=0.9\linewidth]{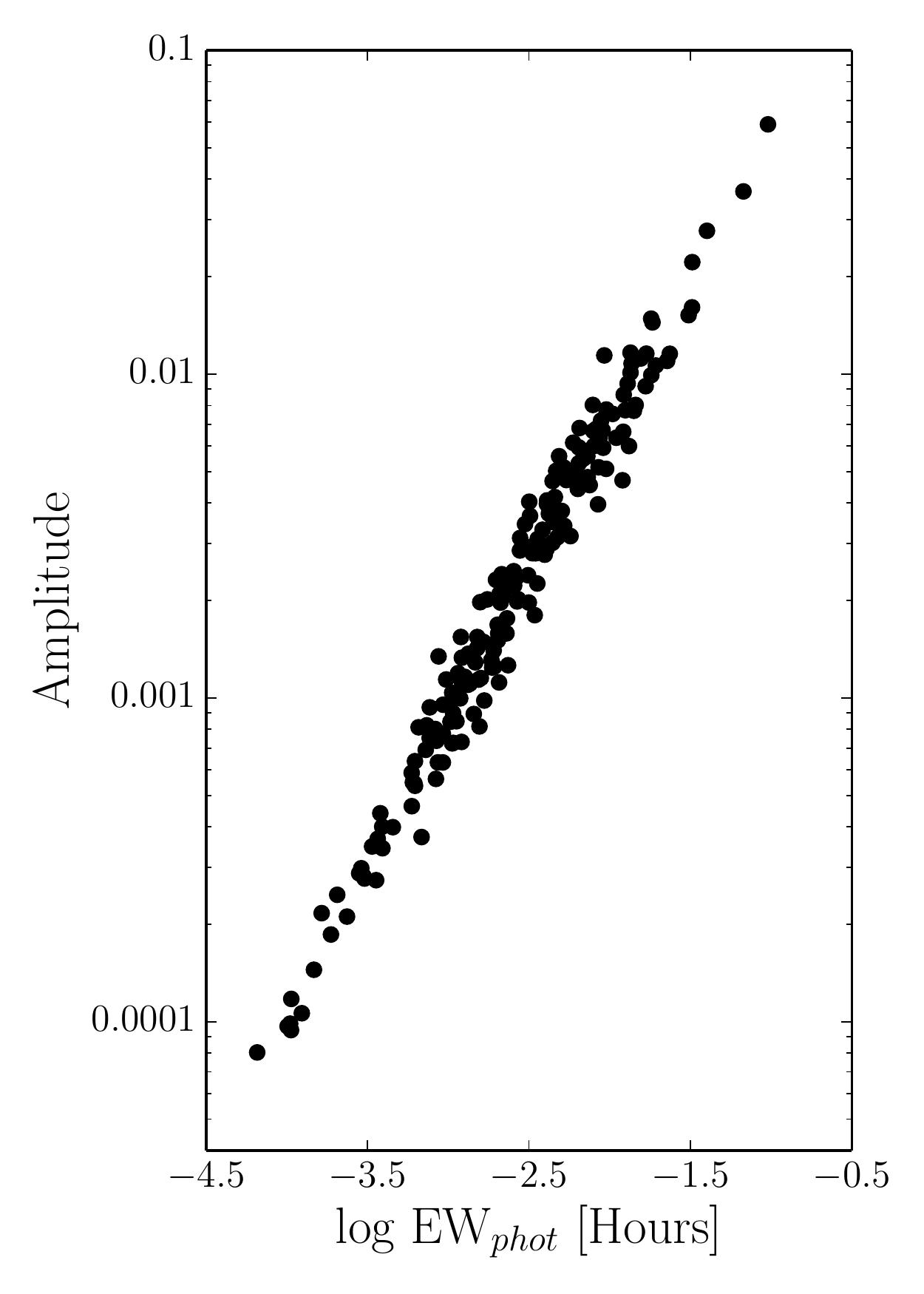}
    \caption{Correlation between flare energy, measured as $EW_\text{phot}$, and amplitude.}
    \label{fig:EWvsAmplitude}
\end{figure}

We explore the energy emitted during the flare ($EW_\text{phot}$) and its duration in Fig. \ref{fig:EWvsDuration}, with energy increasing downwards. As expected, the flare energy increases with increasing duration. The histograms show that the flares tend to be energetic and that the majority of them have a duration between $1.5-2.5 \ \text{h}$, with the lower limit being imposed by our selection criteria. Figure \ref{fig:EWvsAmplitude} shows that the flare amplitude and energy correlate. Again, this is not a surprising behaviour for genuine flares as we expect more energetic flares to have a larger increase in brightness, i.e. amplitude.

\subsection{Fourier Analysis}\label{Sec:Fourier}

Fourier spectra are calculated for each of the A--type stars for which at least one flare has been detected in this work. Based on these spectra, we investigate whether the nature of the quiescent stellar variability, i.e. the variation in the lightcurves excluding the flares, is due to a) pulsations, b) spots and/or c) binarity. 

It is not always straight forward to identify the origin of variability and distinguish between spots and pulsations. We expect that stars of spectral type A are either $\delta$ Sct, $\gamma$ Dor or roAp pulsators. The former two groups usually show coherent multi-periodic variability of the order of $30 \ \text{min}$ to two days. The latter on the other hand, which have strong magnetic fields, have oscillation periods of the order of minutes. 
In the case of spots the time scales should be compatible with rotation periods. Here we can expect two scenarios: (a) short lived spots and possible latitudinal migration as observed in the Sun  will result in a broad peak in the Fourier spectrum; this is because the signal is incoherent. (b) Long-lived magnetic fields on the other hand, as found in Ap/Bp stars can also induce spots.  Due to their long lifetimes these will show sharp coherent peaks in the Fourier spectra with possible harmonics to approximate the non-sinusoidal signal. The latter distribution of peaks could also be explained by ellipsoidal variability in binary systems. Additionally to the Ap/Bp--type stars, HgMn stars also have opacity spots which have been discovered to show some evolution over time \citep[e.g.][]{Kochukhov2007,Briquet2010}.

\begin{figure}
	\includegraphics[width=\linewidth]{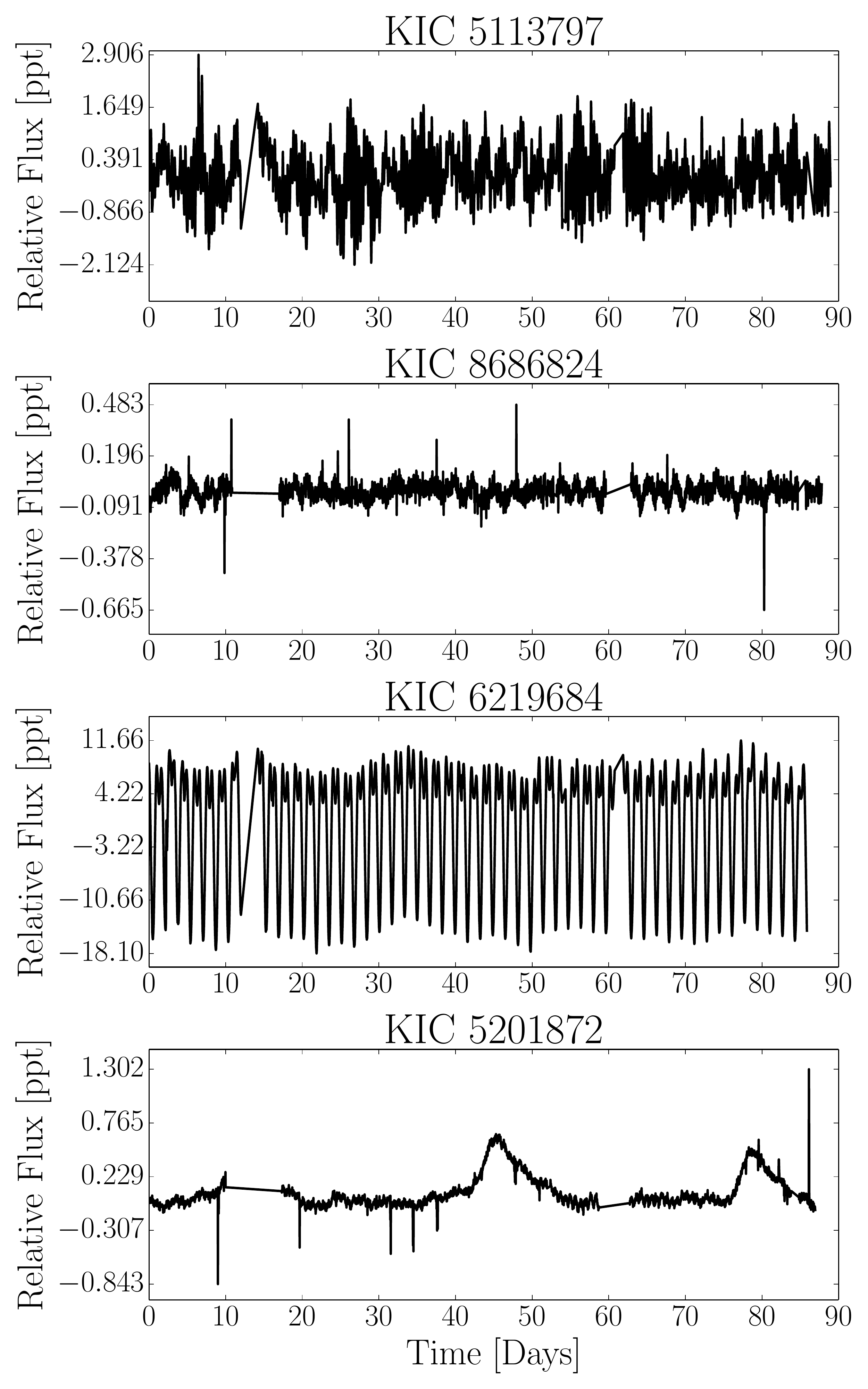}
    \caption{Q2 lightcurves showing examples of stars with pulsations (first panel), short lived spots (second panel), spots and binarity features (third panel) and ellipsoidal variations (fourth panel).}
    \label{fig:Lightcurves}
\end{figure}

\begin{figure}
	\includegraphics[width=\linewidth]{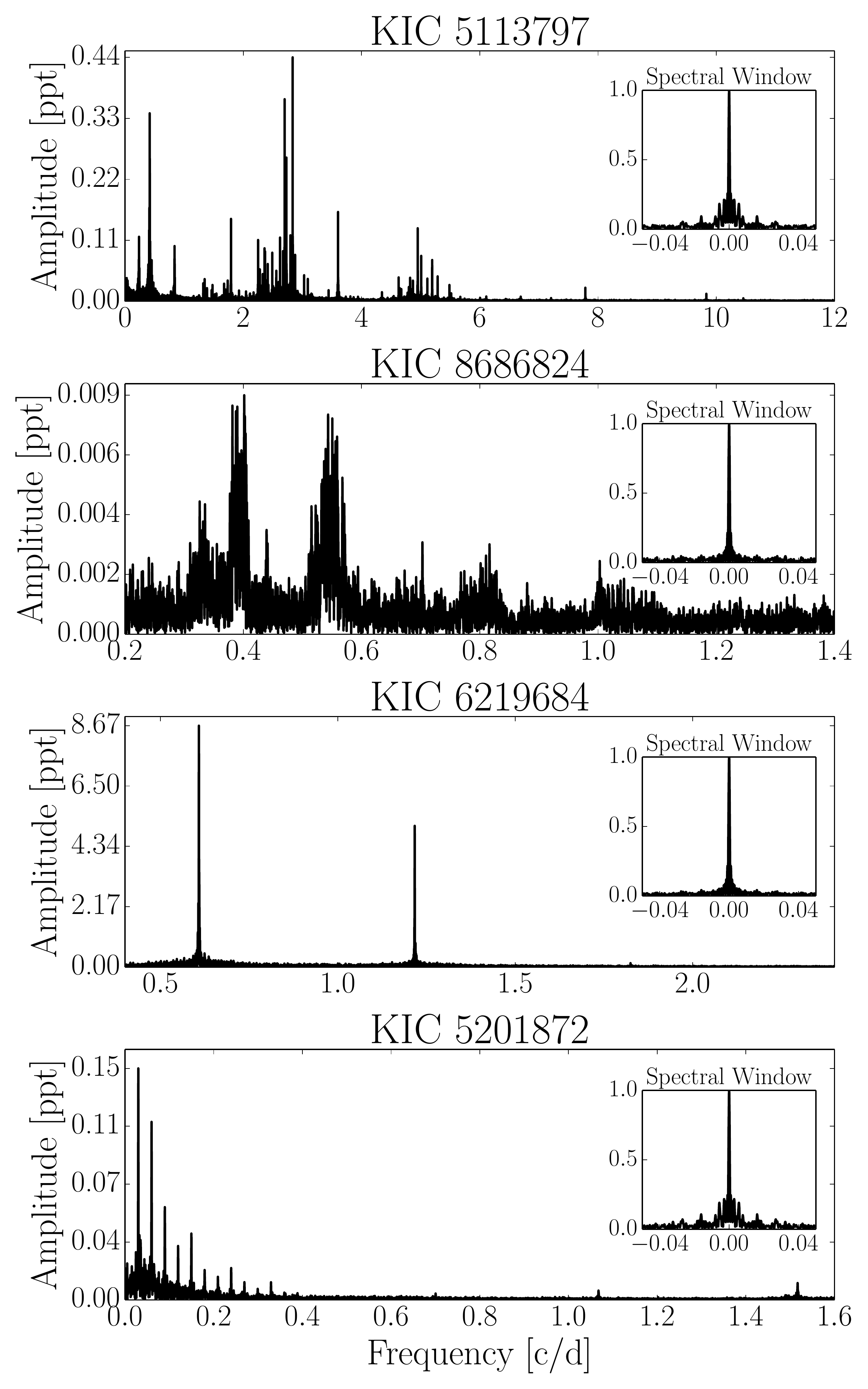}
    \caption{The corresponding Fourier spectra of the lightcurves in Fig. \ref{fig:Lightcurves} including their spectral windows.}
    \label{fig:FourierSpectra}
\end{figure}

Figure \ref{fig:Lightcurves} and \ref{fig:FourierSpectra} show the Q2 light curves of four of our target stars together with their associated Fourier spectra respectively. All four show variability, but are caused by different phenomena, as summarized in Table \ref{Tab:Fourier}. These Fourier spectra have been calculated using \textsc{Period04} \citep{Lenz2005}. The observed frequency peaks are studied and compared to the spectral windows in order to determine their coherency. If harmonics are present in the Fourier spectra, the frequencies and their number of harmonics along with the corresponding periods of the main peaks are given in Table \ref{Tab:Fourier}. In the case of a non-coherent signal, the frequency of the highest amplitude in the non-coherent frequency interval is listed. Non-coherent signals without harmonics are also included, as the harmonics might be buried in the noise. Finally, we identify the variability of each star, listed in the last column of the same Table.  

\begin{table}
	\caption{Results from the Fourier analysis. The first column gives the \emph{KIC} number of the star. $f$ gives the frequency in cycles per day [c/d] of the peaks with harmonics and/or non-coherent shape in the Fourier spectrum. H is the number of harmonics belonging to that frequency and $P$ the corresponding periods of the main peaks. The last column denotes the nature of the frequency spectrum. The binarity nature only refers to results from the frequency analyses and does not include any spectroscopic information.}
	\label{Tab:Fourier}
	\begin{tabular}{cccccccccccccccc}
	\hline
	\text{KIC}	&	\text{$f$}	&	\text{H}		&	\text{$P$}	&	\text{Notes}	\\
	\text{}	& \text{[c/d]}	&	\text{}	&	\text{[d]}	&	\text{}\\
	\hline
  4472809	&	0.424	&	1	&	2.36	&	Spots and binarity\\
  4773133	&	0.076	&	1	&	13.11	&	Spots\\
  5113797	&	0.417	&	1	&	2.40	&	$\gamma$ Dor pulsations + spots\\
  5201872	&	0.0299	&	11	&	33.43	&	Heartbeat star\\
  5213466	&	0.355	&	3	&	2.82	&	Spots and binarity\\
  5559516	&	0.186	&	0	&	5.38	&	Spots\\
  5870686	&	1.334	&	2	&	0.75	&	Spots and binarity\\
  			&	1.395	&	1	&	0.72	&	\\
  5898780	&	0.317	&	1	&	3.16	&	Pulsations + two set spots\\
  			&	0.253	&	1	&	3.95	&	\\
  6219684	&	0.608	&	3	&	1.64	&	Spots and binarity\\
  6451234	&	0.348	&	1	&	2.87	&	Spots, noisy spectrum\\
  7047141	&	0.339	&	1	&	2.95	&	Spots and binarity\\
  7097723	&	0.525	&	2	&	1.90	&	Spots and binarity\\
  7809801	&	0.600	&	0	&	1.67	&	Ellipsoidal variability?\\
  7978512	&	2.153	&	3	&	0.46	&	Spots\\
  8044889	&	0.516	&	3	&	1.94	&	Spots and binarity\\
  8351193	&	0.109	&	0	&	9.16	&	Quiet?\\
  8686824	&	0.402	&	1	&	2.48	&	Spots\\
  			&	0.551	&	1	&	1.816&	Spots\\
  9216367	&	0.308	&	1	&	3.25	&	Spots and binarity\\
  9782810	&	0.077	&	4	&	13.02	&	Spots and binarity\\
  			&	0.906	&	1	&	1.10	&	\\
  10082844	&	0.480	&	3	&	2.08	&	Spots and binarity\\
  10489286	&	0.649	&	1	&	1.54	&	Spots\\
  10817620	&	0.340	&	3	&	2.95	&	Spots and binarity\\
  10971633	&	0.251	&	3	&	3.98	&	Spots and binarity\\
  10974032	&	0.885	&	1	&	1.13	&	Spots and binarity\\
  			&	2.064	&	1	&	0.48	&	\\
  11189959	&	1.216	&	1	&	0.82	&	Spots and binarity\\
  			&	1.633	&	1	&	0.61	&	\\
  11236035	&	0.431	&	3	&	2.32	&	Spots and binarity	\\
  12061741	&	0.340	&	3	&	2.94	&	Spots and binarity\\
  			&	1.852	&	1	&	0.54	&		\\
	\hline
	\end{tabular}
\end{table}

Our analyses reveal that 9 stars have non-coherent signals. Seventeen stars show, in many cases, almost coherent signals with harmonics, see e.g. third panel of Fig. \ref{fig:FourierSpectra}. Many of these peaks are most likely due to long-lived spots with indications of binarity. However, a detailed lightcurve analysis has to be carried out to confirm these binarity features. Furthermore, we note that the lightcurve of KIC 9782810 shows a fairly sharp and periodic increase in brightness followed by a slower decaying phase. The cause of these brightenings may be anything between spots and Cepheid variablity and likewise require further analysis. 

\begin{figure}
	\includegraphics[width=\linewidth]{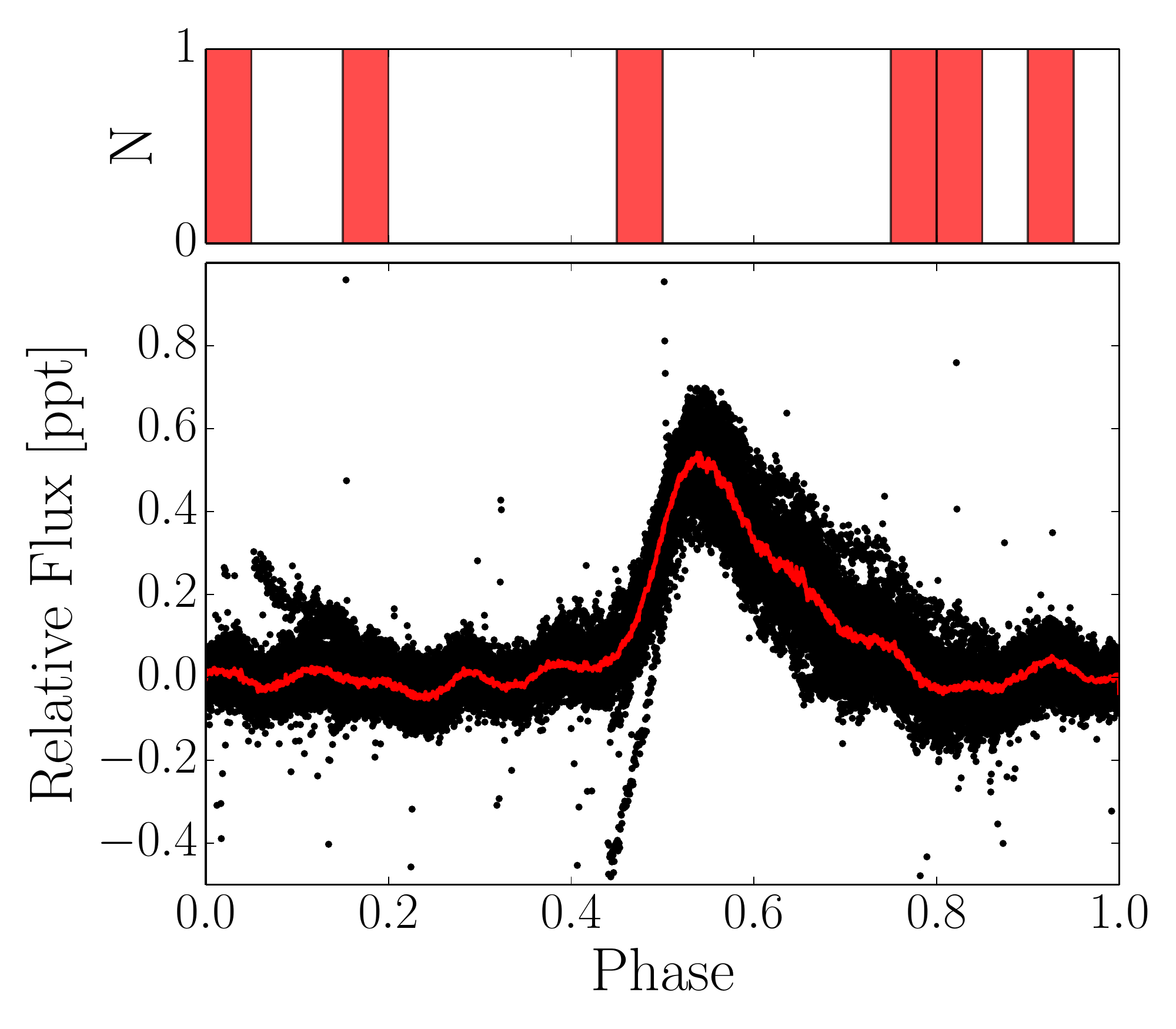}
    \caption{Phase folded lightcurve of the heartbeat star KIC 5201872. The red curve is the running median filtered lightcurve. The upper panel shows the occurrence of flares throughout the orbital period.
    }
    \label{fig:HearbeatPhase}
\end{figure}

The lightcurve of KIC 5201872 seems more exceptional as it undergoes periodic brightenings with durations of the order of days. These periodic events show up as a coherent signal with the largest and a constant amplitude in the Fourier spectrum and has at least 11 harmonics. These kind of brightenings are characteristic for \emph{heartbeat stars}, i.e. stars in eccentric binary systems, and are also known as ellipsoidal variations \citep[e.g.][]{Thompson2012}. They are result of tidal distortion of the system, which becomes strongest at periastron and leads to an increase in brightness. The resulting shape of the flux variation depends on the inclination, angle of periastron and eccentricity of the system \citep[e.g.][]{Thompson2012}. The first detected heartbeat star has a period of $\sim 42 \ \text{d}$ \citep{Welsh2011}, and \citet{Thompson2012} found periods between $\sim 4-20 \ \text{d}$ for 17 different heartbeat stars. Figure \ref{fig:HearbeatPhase} shows the phase folded lightcurve of KIC 5201872 using the period ($P = 33.43 \ \text{d}$), corresponding to the signal with the highest amplitude in the Fourier spectrum. It clearly shows the periodicity of the brightenings. The red curve is obtained by using a running median filter with an interval of 40 data points on the phase folded lightcurve. The upper panel shows the number of flares occurring at a given orbital phase of the binary system. If the observed flares are due to the interaction of magnetic fields between the two stars, it would be expected that the number of flares increases during periastron passage. However, as illustrated in Fig. \ref{fig:HearbeatPhase} this does not seem to be the case.

\section{Pixel Data}\label{Sec:PixelData}

A very important step in this analysis is to determine if the flares originate from the lightcurves of the A--type stars or if nearby stars contaminate the targets. Adopting the terminology of \citet{Coughlin2014}, we refer to such contaminators as \emph{parents} and the contaminated target as the \emph{child}. In order to carry out this investigation, we examine the pixel data and check the entire field-of-view (FOV) covered by the mask.

Photometry is collected by the \emph{Kepler} spacecraft for a number of pixels for each observed target. The collection of pixels for a specific target is called a mask. The full \emph{aperture mask} for a given star consists of the total number of pixels for which data are stored and saved, and contains more than just the pixels covering the target. The \emph{Kepler mask}, is the optimal selection of pixels within the aperture mask which when combined maximise the signal-to-noise \citep[e.g.][]{Bryson2010a}. Both the aperture and \emph{Kepler} mask vary in size from quarter-to-quarter and from star-to-star and depend to a large degree on the \emph{Kepler} magnitude of the target.

\subsection{Neighbourhood}

The first step prior to the pixel data analysis is to study the neighbourhood of the targets, i.e. the FOV in the vicinity of the star that is covered by the \emph{Kepler} mask, in order to look for possible sources of contamination. Using the pixel size \citep[$3\overset{''}{.}98 \times 3\overset{''}{.}98$,][]{VanCleve2009}, we determine the mask dimensions for both the aperture- and \emph{Kepler} mask, see Table \ref{tab:PixelResults}, for the quarter containing the largest flare. This is done for all stars for which we detect at least one flare in this work.

\begin{table*}
\begin{center}
  \small
  \caption{Results from Pixel data analysis. The first column gives the KIC number of the star and \emph{Kp} is the \emph{Kepler} apparent magnitude. The \emph{contamination factor} is a value between 0 and 1 which denotes to the extend of which the lightcurves are expected to be contaminated by light from nearby stars. A contamination factor of zero means that no contamination is expected. Column 4 and 5 gives gives the dimensions of the aperture and \emph{Kepler} mask respectively in units of arcseconds and arcminutes. \textit{Neighbourhood} denotes if there are any other stars in the vicinity (I = isolated, C = close, O = overlapping). The \textit{pixel contamination} column indicates whether the Kepler mask is subject to clear contamination by a nearby star (Y = yes, N = no).}
\begin{tabular}{cccccccccccc}
  \hline
  \text{KIC}		& \text{Kp} 	&	\text{Contamination} &	\text{Mask}	& \text{\emph{Kepler} mask}	&	\text{Neighbourhood}	&	\text{Pixel}\\
    		& \text{[mag]}	&	\text{factor}	&	\text{dimensions}	&	\text{dimensions}	&	&	\text{contamination}\\\hline
  \hline
  4472809	&	14.0	&	0.189	&	$23\overset{''}{.}88 \times 19\overset{''}{.}90$	&	$11\overset{''}{.}94 \times 11\overset{''}{.}94$	&	C/I	&	N\\
  4773133	&	12.1	&	0.084	&	$31\overset{''}{.}84 \times 27\overset{''}{.}86$	&	$19\overset{''}{.}90 \times 19\overset{''}{.}90$	&	C+O/C+O	&	Y\\
  5113797	&	9.1	&	0.06		&	$51\overset{''}{.}74 \times 1\overset{'}{.}06$	&	$23\overset{''}{.}88 \times 55\overset{''}{.}72$	&	C+O/C+O	&	?\\
  5201872	&	9.5	&	0.002	&	$43\overset{''}{.}78 \times 51\overset{''}{.}74$	&	$23\overset{''}{.}88 \times 43\overset{''}{.}78$	&	C+O/C+O	&	N\\
  5213466	&	13.1	&	0.164	&	$27\overset{''}{.}86 \times 23\overset{''}{.}88$	&	$15\overset{''}{.}92 \times 15\overset{''}{.}92$	&	C/C	&	N\\
  5559516	&	8.7	&	0.001	&	$43\overset{''}{.}78 \times 1\overset{'}{.}53$	&	$27\overset{''}{.}86 \times 1\overset{'}{.}33$	&	C+O/C+O	&	?\\
  5870686	&	9.9	&	0.002	&	$43\overset{''}{.}78 \times 39\overset{''}{.}80$	&	$19\overset{''}{.}90 \times 39\overset{''}{.}80$	&	C/C	&	?\\
  5898780	&	12.8	&	0.116	&	$27\overset{''}{.}86 \times 23\overset{''}{.}88$	&	$7\overset{''}{.}96 \times 11\overset{''}{.}94$	&	C+O/C+O	&	Y\\
  6219684	&	12.2	&	0.022	&	$27\overset{''}{.}86 \times 19\overset{''}{.}90$	&	$11\overset{''}{.}94 \times 11\overset{''}{.}94$	&	C/I	&	N\\
  6451234	&	12.8	&	0.085	&	$27\overset{''}{.}86 \times 19\overset{''}{.}90$	&	$19\overset{''}{.}90 \times 15\overset{''}{.}92$	&	C+O/O	&	?\\
  7047141	&	12.7	&	0.148	&	$23\overset{''}{.}88 \times 19\overset{''}{.}90$	&	$11\overset{''}{.}94 \times 11\overset{''}{.}94$	&	C/I	&	N\\
  7097723	&	11.7&	0.027	&	$27\overset{''}{.}86 \times 27\overset{''}{.}86$	&	$19\overset{''}{.}90 \times 23\overset{''}{.}88$	&	O/O	&	Y\\
  7809801	&	14.6	&	0.107	&	$23\overset{''}{.}88 \times 15\overset{''}{.}92$	&	$15\overset{''}{.}92 \times 15\overset{''}{.}92$	&	C/C	&	Y\\
  7978512	&	12.3	&	0.009	&	$27\overset{''}{.}86 \times 23\overset{''}{.}88$	&	$19\overset{''}{.}90 \times 19\overset{''}{.}90$	&	C/I	&	N\\
  8044889	&	12.6	&	0.007	&	$23\overset{''}{.}88 \times 23\overset{''}{.}88$	&	$11\overset{''}{.}94 \times 15\overset{''}{.}92$	&	C/I	&	N\\
  8351193	&	7.6	&	0.024	&	$ 47\overset{''}{.}76\times 4\overset{'}{.}78$	&	$27\overset{''}{.}86 \times 3\overset{'}{.}12$	&	C+O/C+O	&	?\\
  8686824	&	10.7	&	0.008	&	$31\overset{''}{.}84 \times 27\overset{''}{.}86$	&	$19\overset{''}{.}90 \times 27\overset{''}{.}86$	&	C+O/C+O	&	Y\\
  9216367	&	12.1	&	0.106	&	$27\overset{''}{.}86 \times 27\overset{''}{.}86$	&	$15\overset{''}{.}92 \times 19\overset{''}{.}90$	&	C+O/C+O	&	N\\
  9782810	&	9.5	&	0.006	&	$43\overset{''}{.}78 \times 1\overset{'}{.}19$	&	$23\overset{''}{.}88 \times 55\overset{''}{.}72$	&	C+O/C+O	&	?\\
  10082844	&	13.7	&	0.109	&	$23\overset{''}{.}88 \times 19\overset{''}{.}90$	&	$15\overset{''}{.}92 \times 11\overset{''}{.}94$	&	C/I	&	N\\
  10489286	&	11.8	&	0.017	&	$27\overset{''}{.}86 \times 31\overset{''}{.}84$	&	$15\overset{''}{.}92 \times 23\overset{''}{.}88$	&	C+O/O	&	N\\
  10817620	&	14.0	&	0.352	&	$23\overset{''}{.}88 \times 23\overset{''}{.}88$	&	$11\overset{''}{.}94 \times 15\overset{''}{.}92$	&	C/C	&	N\\
  10971633	&	11.5	&	0.001	&	$27\overset{''}{.}86 \times 23\overset{''}{.}88$	&	$19\overset{''}{.}90 \times 15\overset{''}{.}92$	&	O/O	&	N\\
  10974032	&	8.4	&	0.021	&	$51\overset{''}{.}74 \times 3\overset{'}{.}45$	&	$23\overset{''}{.}88 \times 2\overset{'}{.}32$	&	C/C	&	?\\
  11189959	&	8.2	&	0.003	&	$51\overset{''}{.}74 \times 2\overset{'}{.}85$	&	$27\overset{''}{.}86 \times 3\overset{'}{.}25$	&	C/C	&	N\\
  11236035	&	14.6	&	0.04		&	$19\overset{''}{.}90 \times 15\overset{''}{.}92$	&	$7\overset{''}{.}96 \times 7\overset{''}{.}96$	&	I/I	&	N\\
  12061741	&	8.6	&	0.00027	&	$43\overset{''}{.}78 \times 2\overset{'}{.}06$	&	$31\overset{''}{.}84 \times 1\overset{'}{.}59$	&	C+O/C+O	&	?\\
  \hline 
  \end{tabular}
\label{tab:PixelResults}
	 	\end{center}
\end{table*}

\begin{figure}
	\includegraphics[width=\linewidth]{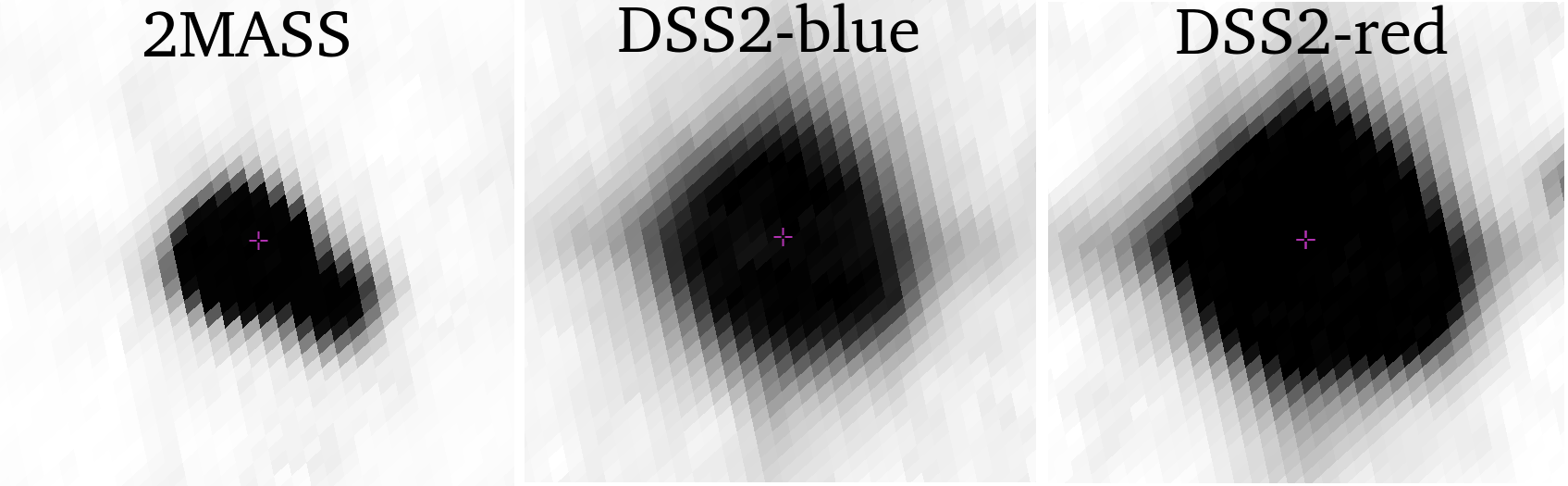}
    \caption{FOV images in different filters for KIC 7097723 with dimensions of $\sim 27\overset{''}{.}86 \times 27\overset{''}{.}86$, matching the mask dimensions listed in column 4 of Table \ref{tab:PixelResults}.}
    \label{fig:KIC7097723_FOV}
\end{figure}

\begin{figure}
	\includegraphics[width=\linewidth]{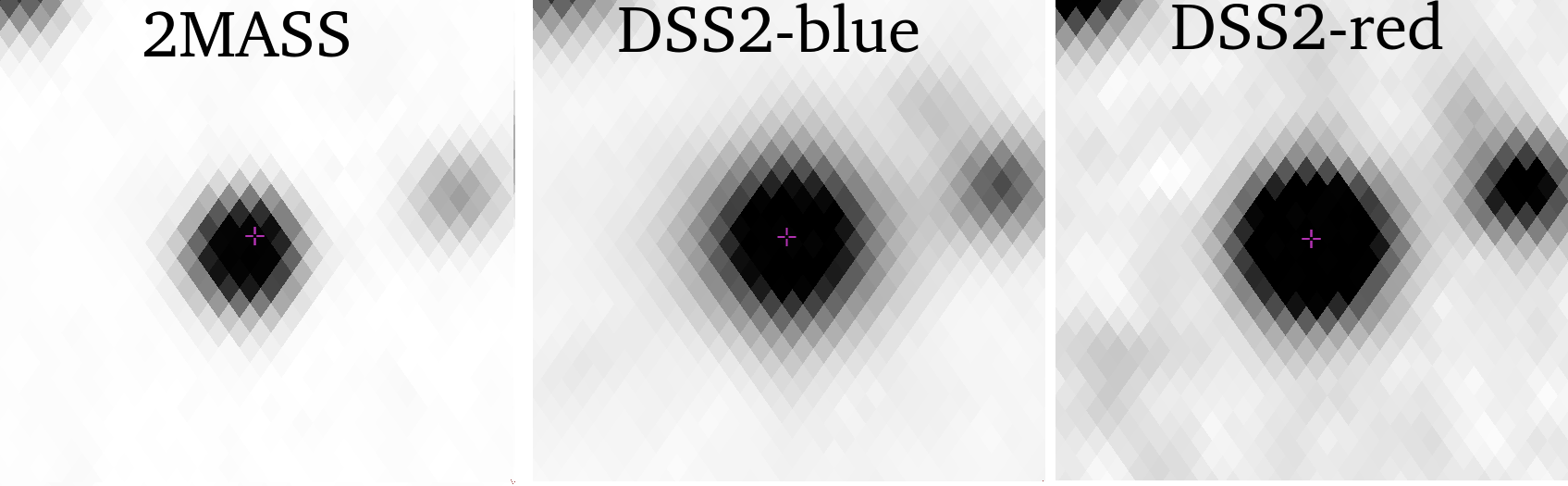}
    \caption{FOV images in different filters for KIC 4472809 with dimensions of $\sim 23\overset{''}{.}88 \times 23\overset{''}{.}88$, matching the maximum width of the mask dimension listed in column 4 of Table \ref{tab:PixelResults}.}
    \label{fig:KIC4472809_FOV}
\end{figure}

We study the FOV within the calculated mask dimensions centred on the target using the interactive \emph{Aladin Sky Atlas} \citep{Bonnarel2000,Boch2014} for three different filters: \textbf{ 1)} the near-infrared Two Micron All Sky Survey (2MASS) images consisting of a combination of images in three different bandpasses $1.25 \ \mu\text{m}$, $1.65 \ \mu\text{m}$ and $1.25 \ \mu\text{m}$ \citep{Skrutskie2006}, \textbf{2)} the second generation Digitized Sky Survey (DSS2) red- and \textbf{3)} the DSS2-blue filter images (STScI/NASA). Figure \ref{fig:KIC7097723_FOV} and \ref{fig:KIC4472809_FOV} show the FOV in these three different filters, covering the maximum height or width of the aperture mask dimensions for KIC 7097723 and KIC 4472809 respectively.

Based on the three FOV images we note the neighbourhood status of each star in Table \ref{tab:PixelResults} under `Neighbourhood'. For this we use an `\textbf{a/b}' syntax, where `\textbf{a}' gives the status within the aperture mask dimensions and `\textbf{b}' the corresponding status within the \emph{Kepler} mask. If no other stars are seen in the FOV in either of the three filters, the star is identified as \emph{isolated, I}. The stars can also be flagged as \emph{close, C,} or  \emph{overlapping, O,} when one or more neighbouring stars overlap with the target in either of the three filters. Figure \ref{fig:KIC4472809_FOV} shows a case where close neighbours are present within the aperture mask, but the star is still isolated in the \emph{Kepler} mask. In comparison, KIC 7097723 is a clear example of a star with an overlapping neighbour, see Fig. \ref{fig:KIC7097723_FOV}. The neighbour is easily seen in the 2MASS FOV image and covers a substantial fraction of the star in the other two filters. 

If we compare these observations with the contamination factor from the KIC catalogue listed in Table \ref{tab:PixelResults} it is immediately evident that the contamination factor is not necessarily a good indicator to identify the flaring star. The contamination factor in the KIC catalogue is determined based on the crowding of stars within an aperture of $83\overset{''}{.}6 \times 83\overset{''}{.}6$\footnote{\url{http://keplerscience.arc.nasa.gov/News.shtml}, Target Crowding and Contamination at MAST,
Dec 21, 2010} square aperture centred on the target, which is larger than the aperture mask dimensions for both KIC 4472809 and KIC 7097723. While a high level of crowding would indeed influence the total flux contribution from the target star, the resulting contamination factor does not provide direct information on whether observed flares originate from a close or overlapping neighbouring star.

Examining the FOV we find that 14 stars out of the 27 stars, for which at least one LC flare has been detected, have overlapping neighbours. Out of these 14 stars, 12 also have close neighbours. One star is isolated both within the aperture- and \emph{Kepler} mask, whereas six stars are isolated within the \emph{Kepler} mask but have close neighbours within the aperture mask. Six stars are neither isolated or have overlapping neighbours, but are found to have close neighbours both within the aperture- and \emph{Kepler} mask. 

\subsection{Pixel contamination}\label{Sec:PixelContamination}

With the neighbourhood status now sorted out, we move on to investigating the pixel data for the most energetic flares for the 27 flaring A--type stars. We only consider cases of \emph{direct contamination} \citep{Coughlin2014}, where light from a parent star is directly included in the target pixels. 

\begin{figure}
	\includegraphics[width=\linewidth]{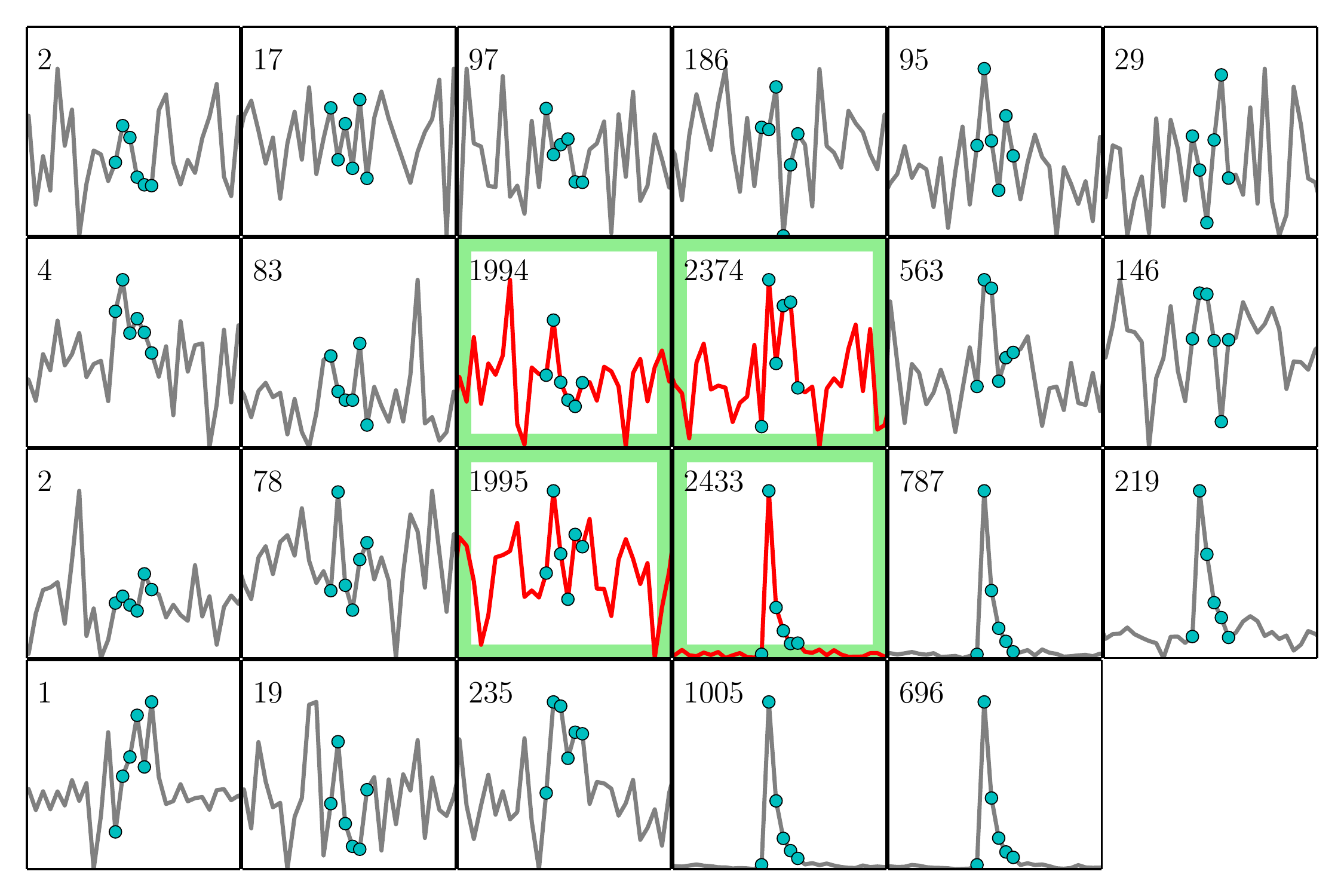}
    \caption{Example of a contaminated star (KIC 7809801). Filled cyan circles identify the flare in the LC lightcurves. Green pixels with red lightcurves denote the pixels belonging to the \emph{Kepler} mask.}
    \label{fig:Contamination}
\end{figure}

\begin{figure}
	\includegraphics[width=\linewidth]{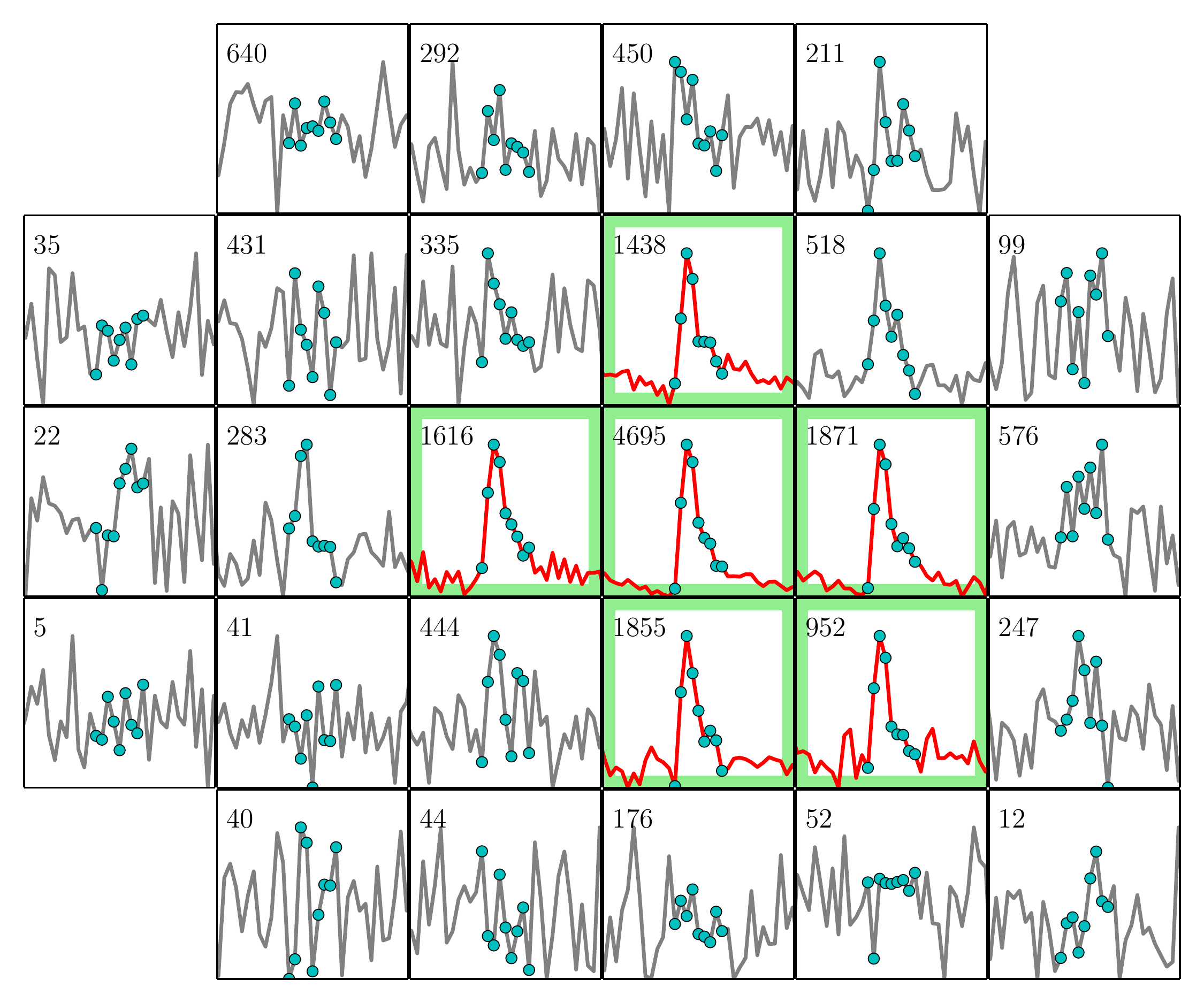}
    \caption{Example of a star without contamination (KIC 4472809).}
    \label{fig:NonContamination}
\end{figure}

For these analyses we plot the aperture mask for the quarters containing the largest flare, see Fig. \ref{fig:Contamination} and \ref{fig:NonContamination}, zooming in on a $\pm 0.25 \ \text{d}$ time interval around the beginning and end of the event. In this interval we correct for the stellar variability as described in Subsection \ref{Sec:FlareDetection}. Pixels belonging to the \emph{Kepler} mask are marked with green borders and red lightcurves. Cyan data points show the expected position of the flare in the lightcurve. The number in the upper left corner of each pixel gives the median of the raw flux divided by the 1-$\sigma$ flux uncertainties, and is an estimate of the S/N in the pixel. If the flare is a result of direct contamination of the \emph{Kepler} mask pixels, we expect that 1) the flare is only seen in a few pixels, and/or 2) the flare amplitude peaks near the border of or outside the \emph{Kepler} mask. If the target is not contaminated, we expect to see the flare in all of the \emph{Kepler} mask pixels.

Figure \ref{fig:Contamination} shows a clear example of a star that is contaminated by a nearby parent source in the bottom right corner of the aperture mask. The flare is only seen in one of the \emph{Kepler} mask pixels and is largest outside of the mask. For KIC 4472809, however, the flare is seen in all of the \emph{Kepler} mask pixels, see Fig. \ref{fig:NonContamination}. Having no overlapping neighbours, this star is marked as not being directly contaminated. 

Results from the pixel data analysis are listed in the last column of Table \ref{tab:PixelResults}. We find evidence of direct contamination for five of the stars, whereas no obvious contamination is found in 14 of the stars. Out of these 14 stars, four have overlapping stars in their neighbourhood which may therefore contribute significantly to the lightcurve of these stars. However, since the flare amplitude is largest around the centre of the target pixels and the overlapping star is faint in comparison, we find that the flare is unlikely to be due to direct contamination. For the remaining eight stars, the analyses are complicated by saturation or high levels of noise which in some cases makes the flares difficult to detect in any of the pixels. At this stage we can neither confirm nor exclude contamination for these stars. We do note, however, that excluding saturated pixels does not remove the flares from the lightcurves.

To summarize, careful analyses of the pixel data reveal that 5 stars are clearly contaminated and therefore can be excluded as the flare source, 8 stars are saturated and a conclusion is impossible and for 14 stars we find no evidence that the flares might originate from a different source.

\section{Spectroscopic analysis}\label{Sec:Spectroscopy}

We have obtained spectra of 22 out of the 33 flaring A--type stars listed by \citet{Balona2012,Balona2013}. The spectra were obtained using the Fibre-fed Echelle Spectrograph (FIES) at the 2.5-m Nordic Optical Telescope (NOT) on La Palma and covers the wavelength range 370-730 nm \citep{Telting2014}. An initial observing round was carried out in June/July 2013 for six stars using the FIES \textsc{Low-Res} fibre with a resolving power $R = 25000$. In October 2013 spectra were obtained once again for six of the brightest A--type stars listed by \citet{Balona2012,Balona2013} using the \textsc{Med-Res} fibre ($R = 46000$). All spectra obtained at a later date were observed using this fibre. In 2014 an observing campaign was carried out to obtain spectra from three epochs for the 10 brightest, flaring A--type stars by \citet{Balona2012,Balona2013}, with the aim to look for possible binary companions. An additional observing campaign was scheduled for October/November 2015 and March 2016 for nine stars, however, due to poor observing conditions the quality of these data is significantly reduced. Furthermore, only six out of the nine targets were observed in 2015 and seven were observed in 2016. For one of our targets observed during this campaign, we obtained further observation in August 2016. 

Combining the data, we end up with spectra for one epoch for four stars, two epochs for five stars and at least three epochs for 13 stars. The target selection was based on the numbers listed in \citet{Balona2012,Balona2013}. All spectra were reduced using \emph{FIEStool}, an open source data reduction software written in \textsc{python} and provided by the staff at NOT\footnote{\url{http://www.not.iac.es/instruments/fies/fiestool/}}. During the reduction process, the spectra were wavelength calibrated using Thorium-Argon wavelength reference spectra obtained either at the beginning or the end of a night of observations. FIEStool employs tasks from the echelle package \textsc{iraf} \citep{Tody1986,Tody1993} for aperture tracing, spectrum extraction and wavelength calibration.

\subsection{Spectral classification and $v \sin i$}\label{Sec:Classification}

For the spectral classification, the merged FIES spectra are used and compared to the Morgan-Keenan (MK) standard stars in \citet{GrayCorbally}. A website is provided in \citet{GrayCorbally} from which the spectra can be downloaded. These digital spectra were originally obtained at the Dark Sky Observatory, are normalised and have a resolution of $1.8 \ \text{\AA}$ per two pixels.

The merged FIES spectra are initially normalised by fitting a polynomial to the continuum. Thereafter the spectral resolution is reduced to match the resolution of the standard stars by convolving the spectra with a Gaussian profile for which the full-width-at-half-maximum corresponds to the desired spectral resolution. Flux values larger than 1.5 for the normalised spectra are removed before the convolution is carried out in order to account for cosmic rays. 

Table \ref{tab:LitPara} lists the spectral classification and stellar parameters known from literature and Table \ref{tab:ClassVsini} displays the results obtained in this work with an uncertainty of $\pm 1$ spectral type. We find good agreement between our results and the spectral classifications performed previously. Furthermore, we confirm that all target stars lie within the activity quiet range from F5-B5.
All targets are A--type stars except for one B9 IV star (KIC 5559516). It is important to note that none of the stars showed spectral features characteristic of Ap/Bp--type stars \citep[e.g. enhanced elements of  \ion{Mn}{ii}, \ion{Eu}{ii}, \ion{Cr}{ii} and \ion{Sr}{ii},][]{GrayCorbally}.

\begin{table*}
\begin{center}
  \small
  \caption{Stellar parameters from literature. The first column gives the KIC number of the star, SpT is the spectral type, T$_\text{eff}$ is the effective temperature, $\log g$ gives the gravity, $v \sin i$ is the projected rotational velocity, [M/H] is the metallicity and $\xi$ is the microturbulence coefficient. Superscripts denote where the listed parameters have been found: $^a$\citet{Tkachenko2013}, $^b$\citet{Floquet1975}, $^c$\citet{Catanzaro2014} and $^d$\citet{McDonald2012}.
  }
\begin{tabular}{ccccccc}
  \hline 
  \text{KIC} 	& \text{SpT} &	\text{T$_\text{eff}$ [K]}	& \text{$\log g$}	&	\text{$v \sin i$ [km s$^{-1}$]}	&	\text{[M/H]}&	\text{$\xi$ [km s$^{-1}$]} \\\hline
  \hline 
  7974841   	&	B9 IV-V$^a$	&	10650 $\pm$ 285$^a$	&	3.87 $\pm$ 0.14$^a$	&	33 $\pm$ 5.0$^a$	&	0.00 $\pm$ 0.13$^a$	&	2.0$^a$		\\
  8351193	&	A0 IV-V $\lambda$ Boo$^a$	&	9980 $\pm$ 250$^a$	&	3.80 $\pm$ 0.15$^a$	&	180.0 $\pm$ 29$^a$	&	-2.35(Fe)$^a$	&	2.0$^a$	\\
  8703413	&	kA2mF0$^b$	&	8000$ \pm$ 150$^f$	&	4.1 $\pm$ 0.1$^c$	&	15 $\pm$	2$^c$	&		&	2.5 $\pm$ 0.2$^c$	\\
  10974032	&	A0IV$^a$	&	9750 $\pm$ 370$^a$	&	3.75 $\pm$ 0.20$^a$	&	270 $\pm$ 32$^a$	&	-0.80(Fe)$^a$	&	2.0$^a$	\\
  11189959	&	- &	9400$^d$	&	-	&	-	&	-	&	-	\\
  11443271	&	-	&	8279$^d$	&	-	&	-	&	-	&	-	\\
  11600717	&	-&	7800$^d$		&	-	&	-	&	-	&	-	\\
  \hline 
    \label{tab:LitPara}
\end{tabular}
 \end{center}
\end{table*}

\begin{table}
\begin{center}
  \small
  \caption{Results from spectral classification and $v \sin i$. The first column gives the \emph{KIC} number of the star, \emph{S/N} is the approximate signal-to-noise ratio in the wavelength range $5788-5862 \ \text{\AA}$, \emph{epoch} is the date of observation (010713 = July 1, 2013), \emph{$v \sin i$} is the projected rotational velocity.
  }
\begin{tabular}{cccccccc}
  \hline 
  \text{KIC}		& \text{S/N} 	&	\text{epoch}	& \text{SpT} 	&	\text{$v \sin i$}\\
  & & & & \text{[km s$^{-1}$]}\\\hline
  \hline 
  3974751	&	$\sim$ 40	&	010713	&	A2 Ve	&	130\\
  4773133	&	$\sim$ 40	&	300613	&	A2 Ve	&	180\\
  5201872	&	$\sim$ 50	&	010713	&	A4 V		&	100\\
  5559516	&	$\sim$ 85	&	060614	&	B9 IV	&	20\\
  5898780	&	$\sim$ 20	&	110316	&	A7 Ve	&	60\\
  6219684	&	$\sim$ 30	&	041115	&	A2 Ve	&	40\\
  7047141	&	$\sim$ 20	&	140316	&	A0 Ve	&	10\\
  7097723	&	$\sim$ 40	&	300613	&	A2 Ve	&	30\\
  7974841	&	$\sim$ 90	&	060614	&	A0 III	&	40\\
  7978512	&	$\sim$ 10	&	130316	&	A1 Ve	&	150\\
  8044889	&	$\sim$ 20	&	180915	&	A5 V		&	30\\
  8351193	&	$\sim$ 85	&	060614	&	A0 V kB8mB7 $\lambda$ Boo	&	190\\
  8367661	&	$\sim$ 80	&	060614	&	A2.5 V	&	150\\
  8703413	&	$\sim$ 90	&	060614	&	kA5hA5mF2 V	&	10\\
  9216367	&	$\sim$ 45	&	300613	&	A3 Ve	&	10\\
  10489286	&	$\sim$ 40	&	010713	&	A5 V		&	80\\
  10971633	&	$\sim$ 45	&	300613	&	A4.5 V	&	20\\
  10974032	&	$\sim$ 95	&	060614	&	A0 V		&	250\\
  11189959	&	$\sim$ 100	&	060614	&	A1 Vb	&	130\\
  11443271	&	$\sim$ 100	&	060614	&	A2 V		&	180\\
  11600717	& 	$\sim$ 90	&	060614	&	A8 V		&	80\\
  12061741	&	$\sim$ 90	&	060614	&	A1 V		& 	180\\
  \hline  
  \end{tabular}
\label{tab:ClassVsini}
	 	\end{center}
\end{table}

We determine approximate values of the projected rotational velocity $v \sin i$ of each star by directly comparing the normalised $R = 25000$ and $R = 46000$ spectra with different synthetic spectra, which have been adjusted to different values of $v \sin i$, until the best match is obtained. For computing the synthetic spectra, we use the stellar spectral synthesis program \textsc{SPECTRUM}\footnote{\url{http://www.appstate.edu/~grayro/spectrum/spectrum.html}} \citep{Gray1994}. We use synthetic spectra representative of the spectral range observed (A0V, A2V, A5V and A7V). To compute those we use the Kurcz model atmospheres\footnote{\url{http://kurucz.harvard.edu/}} \citep{Castelli2004} with $\log g= 4.0$, [M/H]$= 0.00$, $\xi = 2.0 \ \text{km s}^{-1}$ and the following effective temperatures: $T_\text{eff} = 9000 \ \text{K}$ ($\sim$A2), $8250 \ \text{K}$ ($\sim$A5) and $8000 \ \text{K}$ ($\sim$A7). For A0V we use an atmospheric model of Vega with $T_\text{eff} = 9400 \ \text{K}$, $\log g= 3.90$, [M/H]$= -0.50$ and $\xi = 2.0 \ \text{km s}^{-1}$. For all four templates the spectra are adjusted to different values of $v \sin i$ from $10-300 \ \text{km s}^{-1}$ in steps of $10 \ \text{km s}^{-1}$.

The results from this analysis are given in Table \ref{tab:ClassVsini}. We estimate that the uncertainties related to this procedure to be at least $\pm 10 \ \text{km s}^{-1}$ for $v \sin i \leq 100 \ \text{km s}^{-1}$, $\pm 20 \ \text{km s}^{-1}$ for $v \sin i > 100 \ \text{km s}^{-1}$ and $\pm 40 \ \text{km s}^{-1}$ for $v \sin i > 200 \ \text{km s}^{-1}$. 

\subsection{Binarity}\label{Sec:Binarity}

For all obtained spectra, we determined radial velocities ($RV$s) through cross-correlation of each spectral order with synthetic template spectra. The templates were, as in Subsection \ref{Sec:Classification}, calculated using \textsc{SPECTRUM} and Kurucz model atmospheres. Approximate values of parameters from literature, see Table \ref{tab:ModelPara}, are used for computing matching synthetic spectra when available. Otherwise, template spectra for comparable spectral type are computed using the model atmospheres with parameters listed in Table \ref{tab:ModelPara}. The same synthetic spectra were used for determining the $v \sin i$ values in Subsection \ref{Sec:Classification}. 

\begin{table}
\begin{center}
  \small
  \caption{Model parameters of the template spectra. The first four rows list the stars which have specific template spectra computed based on stellar parameters found in literature and listed in Table \ref{tab:LitPara}. The four remaining rows lists the model parameters approximately matching the given spectral types. These templates are used for the stars with no stellar parameters available from literature. $T_\text{eff}$ gives the effective temperature, $\log g$ the gravity, $v \sin i$ the projected rotational velocity, [M/H] the metallicity and $\xi$ the microturbulence for the computed templates.}
\begin{tabular}{cccccc}
  \hline
  \text{KIC} & \text{$T_\text{eff}$}	& \text{$\log g$}	&	\text{$v \sin i$}	&	\text{[M/H]}&	\text{$\xi$}\\
  & \text{[K]}	&	&	\text{[km s$^{-1}$]}	&	&	\text{[km s$^{-1}$]}\\\hline
  \hline
  7974841 	&	10750	&	4.0	&	35	&	0.0	&	2.0  \\	
  8351193	&	10000	&	4.0	&	180	&	0.0	&	2.0	\\
  8703413	&	8000		&	4.0	&	15	&	0.0	&	2.0	\\
  10974032	&	9750		&	4.0	&	280	&	0.0	&	2.0	\\[1ex]
  \hline
  Vega	&	9400		&	3.90	&	varying	&	-0.50	&	2.0\\
  A2 V		&	9000		&	4.0	&	varying	&	0.00		&	2.0\\
  A5 V		&	8250		&	4.0	&	varying	&	0.00		&	2.0\\
  A7 V		&	8000		&	4.0	&	varying	&	0.00		&	2.0\\
  \hline
\end{tabular}
\label{tab:ModelPara}
 \end{center}
\end{table}  

The FIES spectra have a total of 77 spectral orders. Excluding orders with telluric and/or Balmer lines, $RV$s were determined for each remaining spectral order. For stars with high $v \sin i$ values, three additional orders were excluded. Spectral lines in these orders were too shallow to be properly detected and cross-correlated. As a result 29-32 orders were used for the $RV$ determination. Each order is normalised separately by fitting a second order polynomial to the spectrum and are afterwards barycentrically corrected.

The final $RV$ of a star for a given epoch is taken as the average of the $RV$s of the spectral orders. The error on this average is given as 

\begin{equation}
error = \frac{\sigma}{\sqrt{N}},
	\label{eq:RVerror}
\end{equation}

\noindent where $\sigma$ is the standard deviation of the $RV$s and $N$ is the number of orders used in the computation. $RV$ measurements known from literature are listed in Table \ref{tab:LitRV} whereas results from this work are given in Table \ref{tab:RVRes}.

\begin{table}
\begin{center}
  \small
  \caption{Radial Velocities from literature. First column gives the KIC number of the star in question, $RV$ is the measured radial velocity in km s$^{-1}$, and \emph{Reference} lists the papers where the values in the table were found: $^a$\citet{Fehrenbach1990}, $^b$\citet{Fehrenbach1997}, $^c$\citet{Gontcharov2006}, $^d$\citet{Duflot1995}, $^e$\citet{Catanzaro2014}, $^f$\citet{Grenier1999}.}
\begin{tabular}{ccccccc}
  \hline 
  \text{KIC} 	&	\text{$RV$ [km s$^{-1}$]}	&	\text{Reference} \\\hline
  \hline 
  5113797 	&	-3 $\pm$ 5.0&	a\\[1ex]
  5201872	&	-19 $\pm$2.6	&	a\\[1ex]
  5559516	&	-12 $\pm$ 1.9	&	a\\[1ex]
  7974841   	&	13	&	b	\\
  			&	11	&	b\\
  			&	-13	&	b\\
  			&	17	&	b	\\
  			&	13	&	b	\\[1ex]
  8351193	&	-19 $\pm$ 4.5	&	c	\\
  			&	-19	& d	\\[1ex]
  8703413	&	10	&	b\\
  			&	-20	&	b\\
  			&	12	&	b\\
  			&	29	&	b\\
  			&	-32	&	b\\
  			&	-2	&	b\\
  			&	-10	&	b\\
		  	&	-39.68 $\pm$ 1.08	&	e\\[1ex]
  9782810	&	21	&	b\\
  			&	33	&	b\\
  			&	13	&	b\\[1ex]
  11443271	&	-3.9 $\pm$ 4.4	&	f\\
  			&	-9.8 $\pm$ 3.4	&	c\\
  \hline 
\end{tabular}
\label{tab:LitRV}
 \end{center}
\end{table}

\begin{table}
\begin{center}
  \small
  \caption{Results from $RV$ measurements. The first column gives the KIC number of the star, \emph{Epoch} is the time of the year the spectra were obtained, $RV$ is the estimated $RV$ for a given epoch, \emph{SB1} tells if the star is determined to be a single-lined spectroscopic binary based on the estimated $RV$s and the values found in literature given in Table \ref{tab:LitRV} (Y = yes, N = no, Y$^*$ = yes but only because the $RV$s are significantly different from the ones found in literature), and $N_\text{orders}$ gives the number of orders used to obtain the given $RV$s.}
\begin{tabular}{ccccccc}
  \hline 
  \text{KIC} &	\text{Epoch}	&	\text{$RV$ [km s$^{-1}$]}	&  \text{SB1}	&	\text{$N_\text{orders}$} \\\hline
  \hline 
  3974751	&	Jul 2013	&	4.4$\pm$2.9	&	-	&	29	\\[1ex]
  4773133	&	Jun 2013	&	0.8$\pm$5.5	&	-	&	29	\\[1ex]
  5201872	&	Jul 2013	&	-42.6$\pm$1.6	&	Y$^*$	&	32	\\[1ex]
  5559516	&	Jun 2014	&	-2.5$\pm$0.3	&	Y$^*$	&	32	\\
  			&	Aug 2014	&	-2.3$\pm$0.3	&	&	\\
  			&	Oct 2014	&	-2.9$\pm$0.2	&	&	\\[1ex]
  5898780	&	Mar 2016	&	-7.5$\pm$2.2	&	-	&	32\\[1ex]
  6210684	&	Nov 2015	&	-52.2$\pm$1.0	&	N	&	29\\
  			&	Mar 2016	&	-48.1$\pm$1.9	&	\\[1ex]
  7047141	&	Mar 2016	&	-45.0$\pm$0.3	&	?	&	32\\
  			&	Aug 2016	&	-41.4$\pm$0.2	&	\\[1ex]
  7097723	&	Jun 2013	&	62.9$\pm$0.4	&	Y	&	29\\
  			&	Nov 2015	&	0.6$\pm$25.6	&	\\
  			&	Mar 2016	&	44.7$\pm$0.7	&	\\[1ex]
  7974841 	&	Oct 2013	&	-7.0$\pm$0.3	&	Y$^*$	&	32	\\
  			&	Jun 2014	&	-7.5$\pm$0.3	&	&		\\
  			&	Aug 2014	&	-8.1$\pm$0.3	&	&		\\
  			&	Oct 2014	&	-8.0$\pm$0.3	&	&		\\[1ex]
  8044889	&	Oct 2015	&	15.9$\pm$1.0	&	Y	&	32\\
  			&	Mar 2016	&	3.7$\pm$1.1	&	\\[1ex]
  8351193	&	Oct 2013	&	-38.1$\pm$17.9	&	?	&	29	\\
  			&	Jun 2014	&	-11.3$\pm$15.0	&	\\
  			&	Aug 2014	&	-29.6$\pm$8.6	&	\\
  			&	Oct 2014	&	-25.9$\pm$9.8\\[1ex]
  8367661	&	Jun 2014	&	10.5$\pm$2.9	&	N	&	29	\\
  			&	Aug 2014	&	11.7$\pm$3.6	&		&			\\
  			&	Oct 2014	&	11.7$\pm$3.2	&		&			\\[1ex]
  8703413	&	Jun 2014	&	-28.7$\pm$0.1	&	Y	&	32	\\
  			&	Aug 2014	&	5.7$\pm$0.1		&		&		\\
  			&	Oct 2014	&	-41.3$\pm$0.1	&		&		\\[1ex]
  9216367	&	Jun 2013	&	4.0$\pm$0.2	&	Y	&	32	\\
  			&	Nov 2015	&	-1.2$\pm$21.1	&	\\
  			&	Mar 2016	&	-16.6$\pm$0.2	&	\\[1ex]
  10489286	&	Jul	2013	&	-6.6$\pm$1.0	&	N	&	29	\\
  			&	Oct 2015	&	-6.2$\pm$0.8	&	\\
  			&	Mar 2016	&	-7.3$\pm$1.5	&	\\[1ex]
  10971633	&	Jun 2013	&	-13.5$\pm$0.2	&	Y	&	32	\\
  			&	Nov 2015	&	5.5$\pm$1.0	&\\[1ex]
  10974032	&	Oct 2013	&	-5.2$\pm$3.7	&	Y	&	32	\\
  			&	Jun 2014	&	-19.9$\pm$5.8	&	&		\\
  			&	Aug 2014	&	-24.6$\pm$6.9	&	&		\\
  			&	Nov 2014	&	-25.1$\pm$8.1	&	&		\\[1ex]
  11189959	&	Oct 2013	&	-20.9$\pm$1.9	&	Y	&	29	\\
  			&	Jun 2014	&	-16.0$\pm$2.0	&		&		\\
  			&	Aug 2014	&	-9.3$\pm$1.5		&		&		\\
  			&	Oct 2014	&	-4.9$\pm$3.7		&		&		\\[1ex]
  \hline 
\end{tabular}
\label{tab:RVRes}
 \end{center}
\end{table}

\begin{table}
\begin{center}
  \small
  \contcaption{}
\begin{tabular}{ccccccc}
   \hline 
  \text{KIC} &	\text{Epoch}	&	\text{$RV$ [km s$^{-1}$]}	&  \text{SB1}	&	\text{$N_\text{orders}$} \\\hline
   \hline 
  11443271	&	Oct 2013	&	-5.0$\pm$2.5		&	N	&	29	\\
  			&	Jun 2014	&	-5.1$\pm$2.5		&		&		\\
  			&	Aug 2014	&	-5.2$\pm$3.1&		&		\\
  			&	Nov 2014	&	-8.9$\pm$2.8	&		&		\\[1ex]
  11600717	&	Oct 2013	&	-18.5$\pm$0.8	&	N	&	32	\\
  			&	Jun 2014	&	-18.8$\pm$0.5	&		&		\\
  			&	Aug 2014	&	-18.5$\pm$0.6	&		&		\\[1ex]
  12061741	&	Jun 2014	&	-9.5$\pm$3.8	&	Y	&	29	\\
  			&	Aug 2014	&	0.1$\pm$5.0		&		&		\\
  			&	Nov 2014	&	2.2$\pm$3.9		&		&		\\[1ex]
   \hline 
\end{tabular}
\label{tab:RVRes2}
 \end{center}
\end{table}

Combining the determined $RV$s with measurements from literature we have $RV$s for $\geq 2$ epochs for 19 stars. Out of these we find 11 single-line-spectroscopic-binaries (SB1s), five single stars and three unknown. These measurements confirm the binarity of the suspected heartbeat star KIC 5201872. Furthermore, all of the SB1s are listed as showing binarity features in Table \ref{Tab:Fourier}. Among the three unknown, one is a $\lambda$ Boo star (KIC 8351193) which are known for having a low metallicity for their spectral type, which in return corresponds to few lines making RV measurements complicated or impossible. Coupled with a high $v \sin i$, this results in large uncertainties in the determined $RV$s. The second star with unknown binarity status is KIC 7978512. It likewise has a high $v \sin i$ and low data quality. For the same reason the determined $RV$s for KIC 7978512 are not listed in Table \ref{tab:RVRes}. For KIC 7047414 the measured $RV$ shift is too small to clearly state that the star is in a binary system.

\subsection{Circumstellar vs Interstellar material}\label{Sec:CSvsIS}

\begin{figure}
	\includegraphics[width=\linewidth]{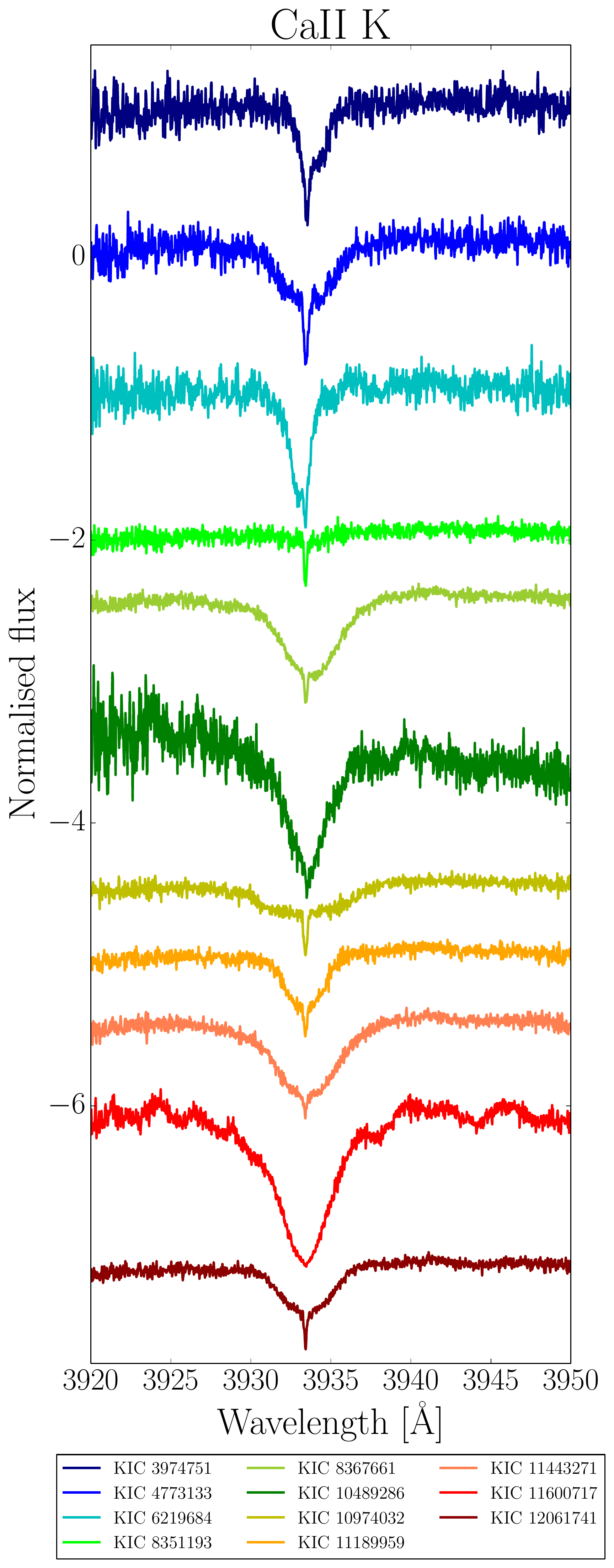}
    \caption{Examples of additional, narrow absorption features seen in the center of the \ion{Ca}{ii} K line.}
    \label{fig:GH}
\end{figure}

\begin{figure}
	\includegraphics[width=\linewidth]{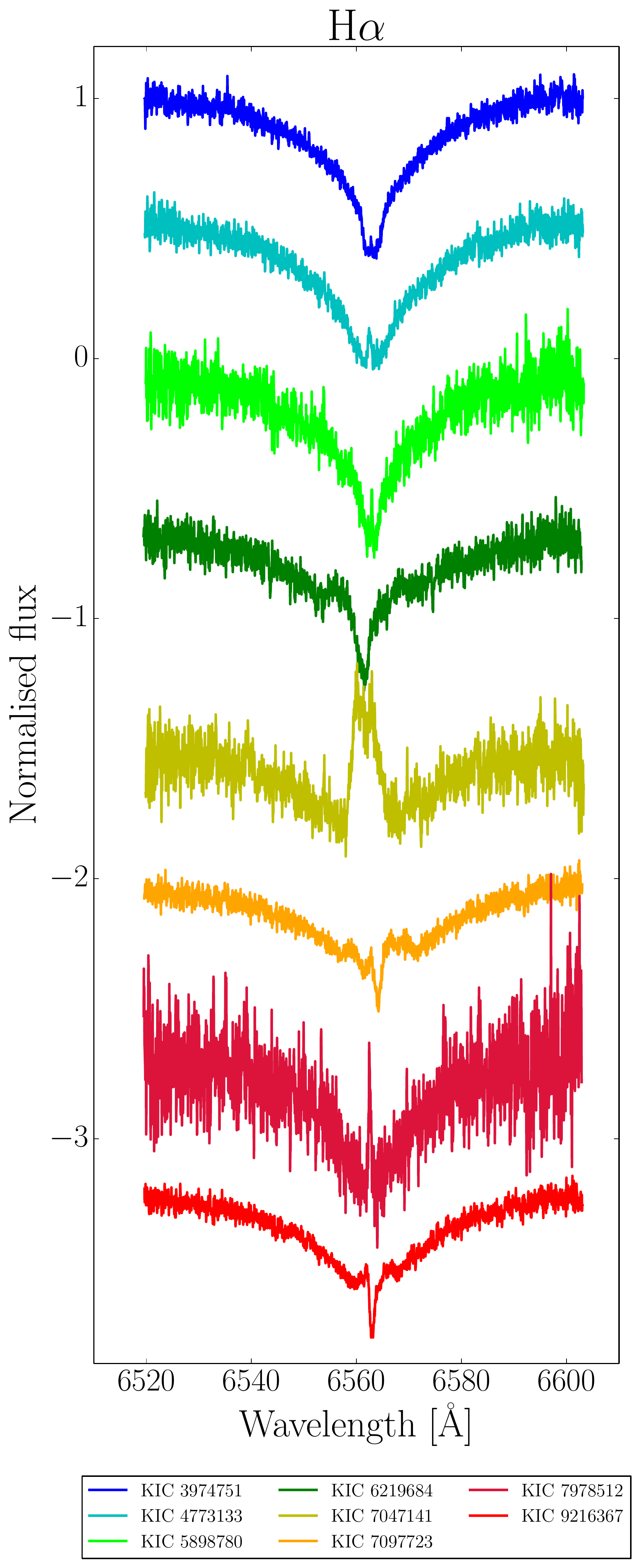}
    \caption{Observed H$\alpha$ emission.}
    \label{fig:Halpha}
\end{figure}

While analysing the FIES spectra we noticed the presence of additional absorption at or near the center of the \ion{Ca}{ii} K line in 11 of the stars, see Fig. \ref{fig:GH}, while the presence of such additional absorption is uncertain for an additional six stars. 14 stars show  absorption in the \ion{NaD}{} lines, and maybe also in another six stars. At least nine stars show both narrow absorption features in both \ion{Ca}{ii} K and \ion{NaD}{} lines. Furthermore, we find emission in H$\alpha$ for eight of the stars, see Fig. \ref{fig:Halpha}, out of which three also show clear \ion{Ca}{ii} K line absorption

Additional absorption in the \ion{Ca}{ii} K and \ion{NaD}{} lines can be caused by interstellar material \citep[e.g.][]{Welsh2010} or by circumstellar gas as in the case of $\beta$ Pictoris \citep[e.g.][]{Slettebak1975,Slettebak1982,Hobbs1985}. An immediate way to distinguish between the two scenarios is by determining the $RV$s of the narrow absorption features. If the $RV$s are similar to the stellar RVs, the observed features are likely to be circumstellar. Otherwise the more likely explanation is that they are interstellar. 

We determine the barycentric velocities of the narrow absorption features by fitting them as cloud components, following the procedure described in \citet{Vallerga1993}. To carry out the fitting we use the following oscillator strengths of the \ion{Ca}{ii} K and \ion{NaD}{} lines respectively: \ion{Ca}{ii} K) 0.688 \citep{Cardelli1986}, \ion{NaD}{i}) 0.3199 and \ion{NaD}{ii}) 0.6405 \citep{Steck2008}. Before modeling the \ion{NaD}{} lines, we correct the spectra for atmospheric absorption lines. This is done by computing telluric line model spectra for each star and observing time using \textsc{TAPAS}\footnote{\url{http://ether.ipsl.jussieu.fr/tapas/}} (Transmissions Atmosphériques Personnalisées Pour l'AStronomie), which is an online modelling tool used to compute atmospheric transmission line spectra \citep{Bertaux2014,Bertaux2014b}. After correcting for the telluric lines, both of the \ion{NaD}{} lines are fitted simultaneously, requiring that the same cloud component velocity is found for both lines.

We find for all cloud components that the barycentric velocities for the \ion{Ca}{ii} K line lie within $-9.4$ to $-20.2 \ \text{km s}^{-1}$ and from $-6.3$ to $-27.2 \ \text{km s}^{-1}$ for the \ion{NaD}{} lines. The \ion{Ca}{ii} K line velocities are comparable to the \ion{NaD}{} velocities for at least one of the cloud components for all stars. When compared to the stellar $RV$s listed in Table \ref{tab:RVRes}, however, none of the cloud component velocities match the stellar $RV$s. The only exception is KIC 11600717, which we have found not to be flaring. Furthermore, for the binary stars none of the cloud component velocities varies for the different epochs even though the corresponding stellar $RV$s do. Therefore, we conclude that the observed narrow absorption features are more likely caused by interstellar- rather than circumstellar gas. We note that it is still possible that circumstellar features may be hidden within the interstellar absorption lines, but that the spectral resolution is too low to resolve those. For large differences between the cloud component and stellar $RV$s this is, however, unlikely.

For the eight stars showing H$\alpha$ emission two are in binary systems and one is a single star. For the remaining five stars we do not have sufficient spectral data available to determine their binary nature. Among the eight stars, seven of them only show emission in H$\alpha$ whereas KIC 7047414, which also has the strongest H$\alpha$ emission as seen in Fig. \ref{fig:Halpha}, has multiple Balmer lines in emission. Furthermore, the Balmer line emission has decreased between the two epochs (from March 2016 to August 2016) and is only observed in the H$\alpha$ line for August 2016. KIC 7978512 also shows variation in the H$\alpha$ line emission from one epoch to the next.

Possible explanations for the observed emission are a) magnetic activity, b) a circumstellar disk or c) mass-transferring binary systems. The fact that the narrow absorption features in \ion{Ca}{ii} K and \ion{NaD}{} lines are interstellar does not exclude that the H$\alpha$ emission can be caused by circumstellar material. Infra-red measurements could, however, confirm it. 

KIC 9216367 shows possible, weak emission in the center of the \ion{Ca}{ii} K line. With the available S/N it is, however, uncertain whether or not this feature is genuinely there. Emission in \ion{Ca}{ii} K is an indicator of magnetic activity. KIC 9216367 is a binary, therefore the observed emission may originate entirely from the companion star. For the remaining seven stars no emission is seen in the \ion{Ca}{ii} K line. The stars showing emission in H$\alpha$ are marked with an `e' in Table \ref{tab:ClassVsini}.

\section{Discussion}\label{Sec:Discussion} 

\begin{figure}
	\includegraphics[width=\linewidth]{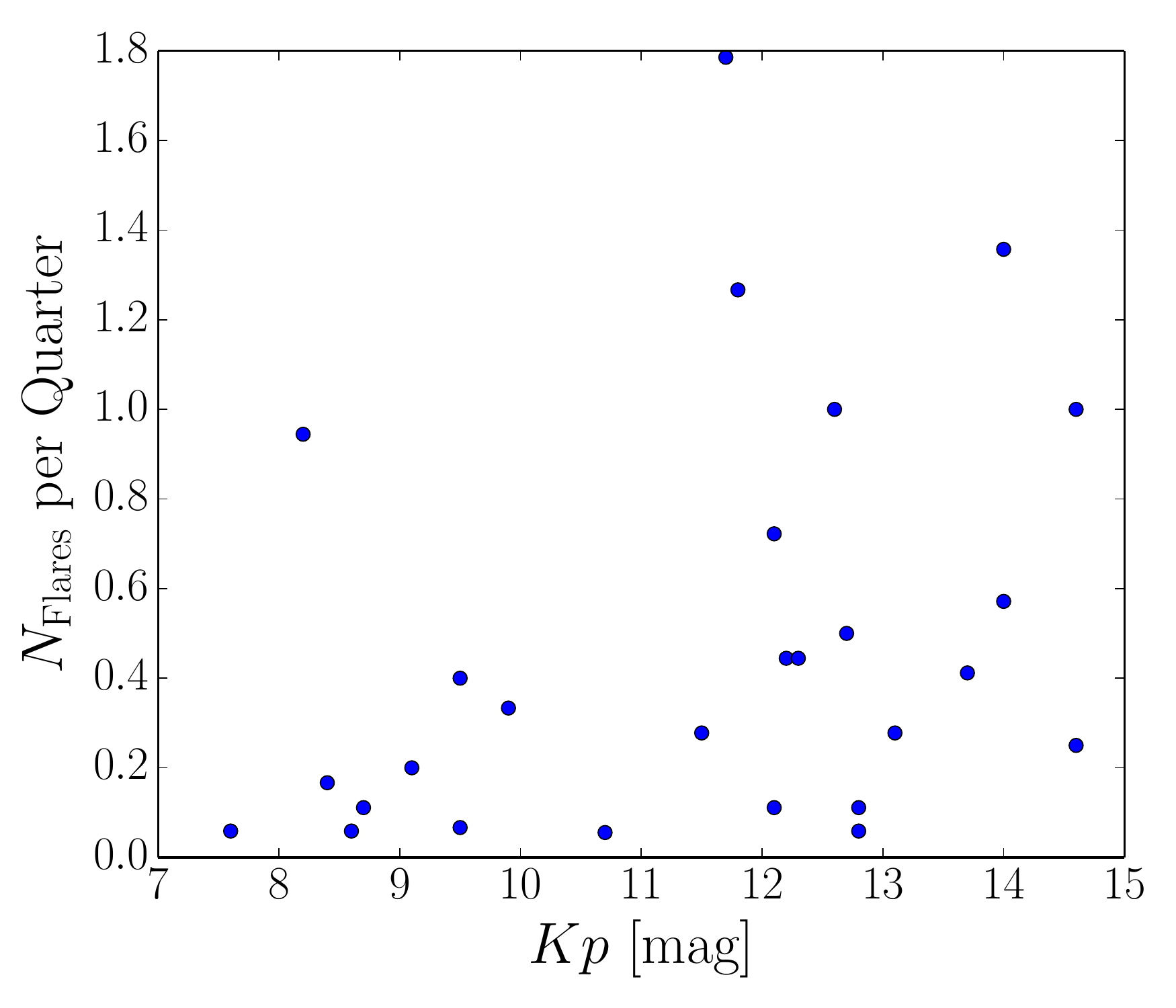}
    \caption{Number of detected flares per LC quarter as a function of \emph{Kepler} magnitude. Each data point corresponds to a flaring A--type star.}
    \label{fig:KpvsNFlares}
\end{figure}

\begin{figure}
	\includegraphics[width=\linewidth]{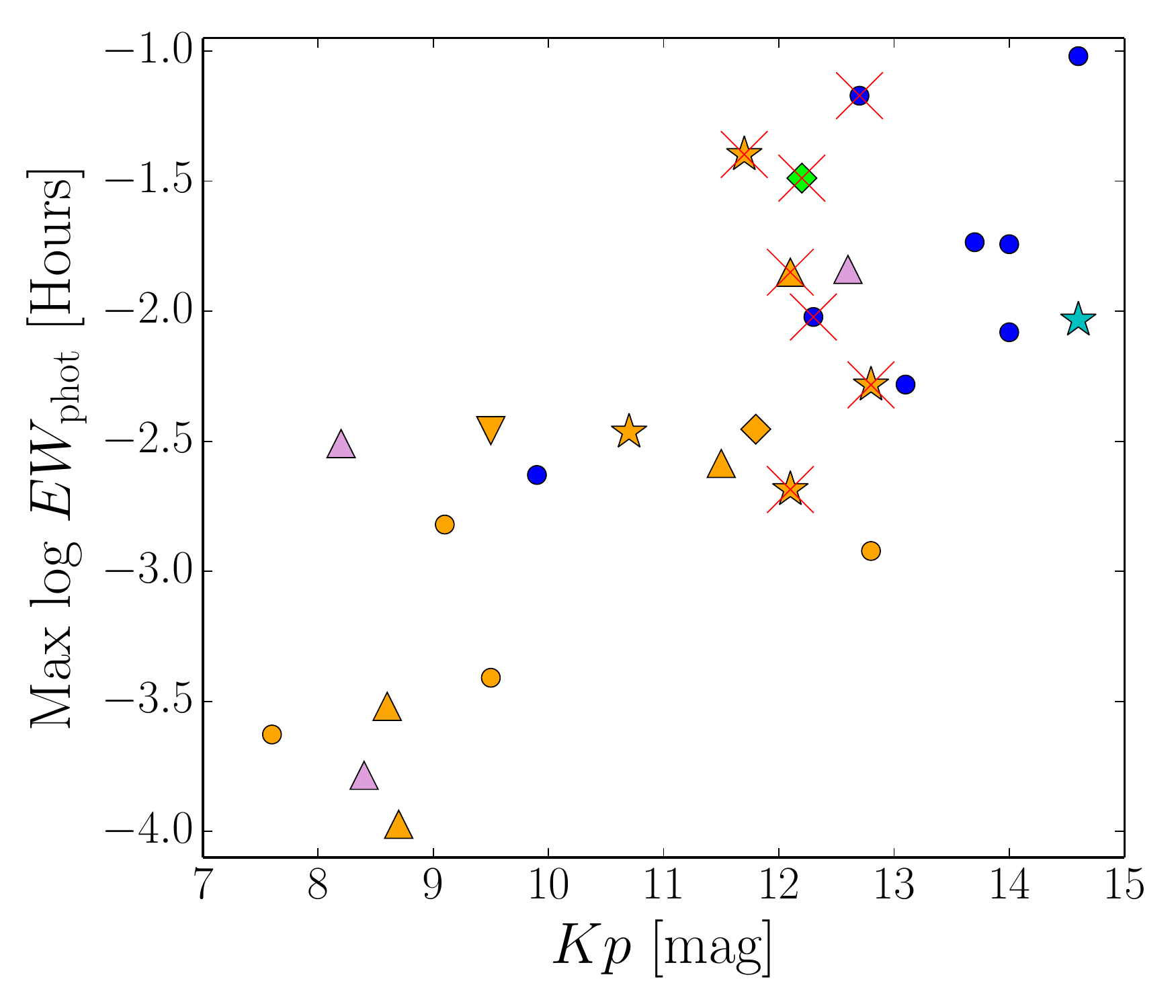}
    \caption{Maximum detected flare energy as a function of \emph{Kepler} magnitude. Each data point corresponds to a flaring A--type star. \emph{Orange data points}: stars with overlapping neighbours in the FOV. \emph{Stars}: contaminated stars. \emph{Triangles}: SB1s. \emph{Inverse triangle}: heartbeat star. \emph{Diamonds}: single stars. \emph{Red crosses}: H$\alpha$ emission. \emph{Blue filled circles}: No special remarks.}
    \label{fig:KpvslogEWmax}
\end{figure}

Studying the relation between the number of detected flares, maximum flare energies and the \emph{Kepler} magnitudes, $Kp$, of the stars can give an indication of whether or not the observed events are due to contamination. Each data point in Fig. \ref{fig:KpvsNFlares} represents a flaring A--type star and shows the relation between the number of detected flares per quarter and the $Kp$. As illustrated, there is a tendency for the number of flares per quarter to increase as the \emph{Kepler} magnitude increases. This correlation becomes even stronger when the maximum detected flare energy of each star is plotted against their $Kp$ as shown in Fig. \ref{fig:KpvslogEWmax}. 

In Fig. \ref{fig:KpvslogEWmax} we mark the 14 stars with detected overlapping neighbours with orange data points. As seen in the figure, most of the bright stars are marked as such. Light from an overlapping neighbouring star will dilute the light from the flaring star. As a consequence the flare will seem less energetic. The same effect is expected for stars in binary systems, as further discussed below. If the light contribution from the diluting stars is removed, then both the \emph{Kepler} magnitude and flare energy would increase. This would cause the data points to move towards the upper right corner in the figure and make the correlation disappear. Therefore, we suspect that the observed correlation between the brightness (\textit{Kp}) and flare energy ($EW_\text{phot}$) is an observational artifact.

The five stars with clear evidence of direct contamination from neighbouring stars are marked by stars in Fig. \ref{fig:KpvslogEWmax}. 14 stars showed no such clear evidence of direct contamination, however, in order to completely rule this out as a possible explanation for the observed flaring, other sources of contamination, such as those discussed in \citet{Coughlin2014}, should be investigated. \citet{Balona2013} did consider the SC pixel data of the two stars with the largest flares detected, which coincidently turned out to be isolated. Based on these two stars \citet{Balona2013} concludes that none of the observed flares can be caused by contamination. As seen in this work, a star-by-star analysis reveals a different result. For completion, we also mark the stars with H$\alpha$ emission (red crosses), single stars (diamonds), the heartbeat star (downward-pointing triangle) and the remaining 10 SB1s (triangles) in Fig. \ref{fig:KpvslogEWmax}. 

\citet{Balona2012,Balona2013} argues that the observed flaring cannot be caused by a cool, unresolved companion. The key argument is that if the flares originated from a cool companion, these would have to be a 100 times more energetic than flares normally detected in M-- and K--dwarfs \citep{Walkowicz2011}. A release of such high energies during flaring is not easily explained based on the present-day models, leading \citet{Balona2012,Balona2013} to conclude the flare cannot originate from cool companions. For these computations \citet{Balona2012} compares the average of the most energetic flare in all A-type stars to the average of all the median flare energies observed in M and K dwarfs derived by \citet{Walkowicz2011}. Here we adopt a slightly different approach by considering the SB1s independently and analyse the energies and amplitudes of the most energetic flares under the assumption that the flare originates from the cool companion. We estimate the undiluted flare energies, $EW_\text{phot,Companion}$, and amplitudes, $A_\text{Companion}$, by correcting for the brightness difference between the A--type primary star with an absolute magnitude $M_\text{V,1}$ and the assumed cool companion with an absolute magnitude of $M_\text{V,2}$ using the relations

\begin{align*}
x &= M_{V,2} - M_{V,1}\\
v_b &= 10^{0.4 \cdot x}\\
EW_\text{phot,Companion} &= \int v_b \cdot \frac{F_\text{rel} - F_\text{q}}{F_\text{q}} dt\\
A_\text{Companion} &= v_b \cdot \frac{\Delta F}{F} =  v_b \cdot \left(\frac{F_\text{rel} - F_\text{q}}{F_\text{q}}\right)_\text{peak}.
\end{align*}

Based on values given in \citet{Allen1973} we use $M_\text{V,1} = 2.0$ for stars corresponding approximately to A5 main-sequence stars (KIC 5201872, KIC 8044889, KIC 10971633) and $M_\text{V,1} = 0.7$ for $\sim$A0 V type stars (KIC 5559516, KIC 10974032, KIC 11189959, KIC 12061741). For KIC 9216367 we use the absolute visual magnitude reported for the A3 V standard star Fomalhaut \citep[$M_\text{V,1} = 1.722$, ][]{Mamajek2012}.
The choice of absolute magnitude is based on the assigned spectral type in Table \ref{tab:ClassVsini}. $M_\text{V,2}$ is varied between 4.4 and 17 mag. 

\begin{figure}
	\includegraphics[width=\linewidth]{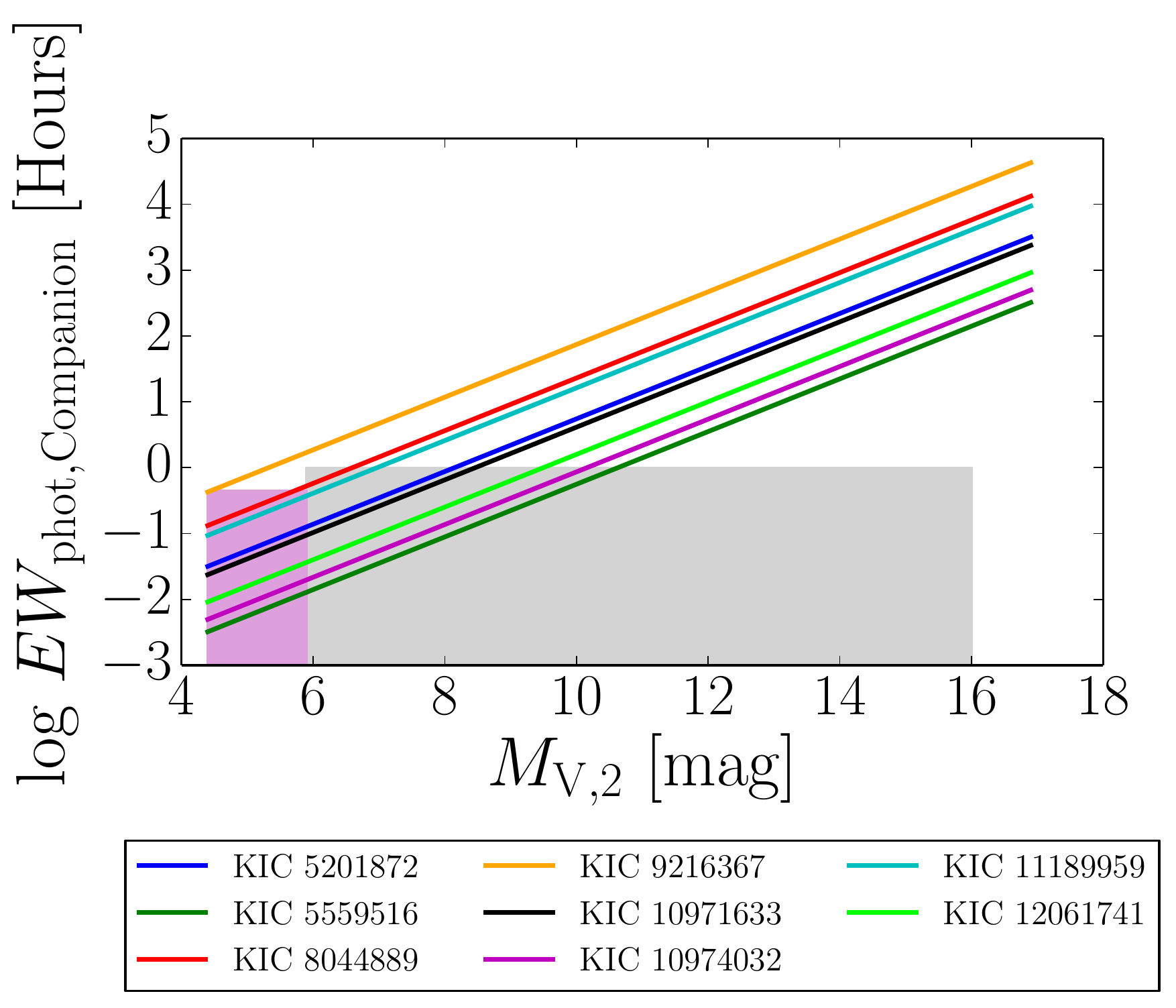}
    \caption{The variation of $EW_{\text{phot,Companion}}$ as a function of the absolute visual magnitude, $M_{V,2}$, of a possible unresolved companion. The curves show how intense the largest detected flares for the identified binaries would be if they originated from a cooler companion. The vertical boundaries of the grey, shaded region marks the absolute visual magnitudes corresponding to main-sequence stars between K0 and M8. The upper boundary gives the energy of the largest white light flare detected by \citet{Walkowicz2011}. The purple box marks the absolute magnitude region of main-sequence G--type stars, with the upper boundary corresponding to the largest superflare energy found in the G--type superflare star KIC 9216367. If the curves cross either of the boxes, then the observed flare could originate from a cool companion.}
    \label{fig:DillutionEffectsI}
\end{figure}

\begin{figure}
	\includegraphics[width=\linewidth]{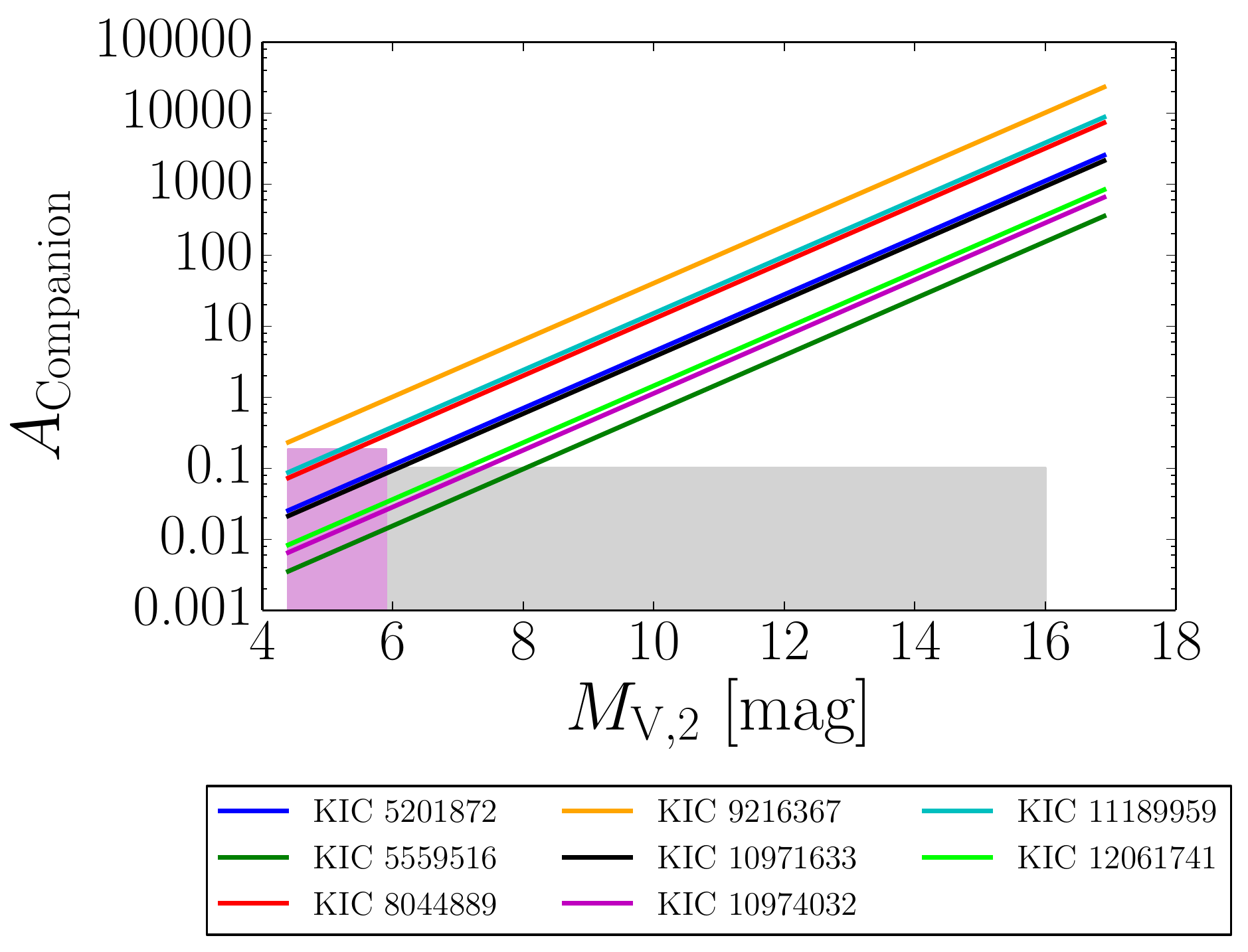}
    \caption{Same as for Fig. \ref{fig:DillutionEffectsI} but for flare amplitudes instead of energies. The curve for KIC 9216367 overlaps nearly perfectly with the curve for KIC 11189959 and is therefore difficult to see in the figure.}
    \label{fig:DillutionEffectsAmplitude}
\end{figure}

Figures \ref{fig:DillutionEffectsI} and \ref{fig:DillutionEffectsAmplitude} show the estimated flare energies and amplitudes as a function of $M_\text{V,2}$. The grey shaded regions cover the absolute magnitude range of K0 to M8 dwarfs \citep{Allen1973} and the maximum flare energies and amplitudes found for the sample of K-- and M--dwarfs in \citet{Walkowicz2011}. Here the median flare energy is determined for each of the flaring late--type stars. Therefore the upper boundary of the grey shaded region gives the maximum of all the median flares in the sample. The purple shaded region represents the absolute magnitude range of G0 to K0 dwarfs \citep{Allen1973}. In order to obtain an upper boundary for the flare energy and amplitude in this magnitude range, we carried out a flare detection and parameter determination for one of the G--type superflare stars (KIC 9603367) listed in \citet{Shibayama2013}. The upper boundary of the purple box corresponds to the energy and amplitude of the most energetic flare found in the LC \emph{Kepler} data of this star.

As seen in Fig. \ref{fig:DillutionEffectsI}, all of the computed $EW_\text{phot,Companion}$ vs $M_\text{V,2}$ curves cross either the purple and/or the grey box at some point. The same is the case when the amplitudes ($A_\text{Companion}$) vs $M_\text{V,2}$ curves are considered in Fig. \ref{fig:DillutionEffectsAmplitude}.

For this analysis we are in fact comparing the maximum detected flare energies of the A--type stars with the maximum of the median flare energies of the K-- and M--dwarfs. If instead the maximum energy of the largest flares of the K-- and M--dwarfs were known and used for this comparison, the upper boundary of the grey box would move upwards in both Fig. \ref{fig:DillutionEffectsI} and \ref{fig:DillutionEffectsAmplitude}, which would cause the curves to cross the boxes at an even higher $M_\text{V,2}$, i.e. lower companion mass. Furthermore, we only computed the flare energies of one G--type superflare star and used this for comparison for the purple box. It is quite possible that other superflare stars have flares with larger energies, which if included would likewise cause the upper limit of the purple box to move upwards. Contrary to \citet{Balona2012,Balona2013}, we therefore conclude that the flares in the detected binary systems can in fact originate from a cool companion.

In \citet{Balona2012}'s calculation of the average flare energies, it has been assumed that all the flares originate from a cool companion. However, as seen in this work the flares can also be caused by contamination from nearby stars. In these cases it is the apparent rather than the absolute magnitudes which need to be compared when considering dilution from the combined light of the systems. By including contaminated stars, this can therefore result in much larger flare intensities and does not represent a true average of the flare intensities. In comparison, when considering the SB1s separately we find that a cool companion may well explain the observed flares as illustrated in Fig. \ref{fig:DillutionEffectsI} and \ref{fig:DillutionEffectsAmplitude}. We therefore emphasise that when investigating the origin of flares in A--type stars it is crucial to study all possible explanations and check each star separately. 

\begin{table}
	\caption{Overview of the possible explanations found in this work for the observed flaring in the A--type stars listed in \citet{Balona2012,Balona2013}. Stars with nothing listed and/or question marks require further investigation.}
	\label{Tab:Explanations}
	\begin{center}
	\begin{tabular}{cc}
	\hline
	\text{KIC}	& \text{Possible explanation(s)}\\\hline
	\hline
	3974751	&	Instrumental\\[0.5ex]
  4472809	&	Binary?\\[0.5ex]
  4773133	&	Contamination\\[0.5ex]
  5113797	&	Overlapping neighbour?\\[0.5ex]
  5201872	&	Cool companion\\
  			&	Overlapping neighbour?\\[0.5ex]
  5213466	&	Binary?\\[0.5ex]
  5559516	&	Cool companion\\
  			&	Overlapping neighbour?\\[0.5ex]
  5870686	&	Binary?\\[0.5ex]
  5898780	&	Contamination\\[0.5ex]
  6219684	&	-\\[0.5ex]
  6451234	&	Overlapping neighbour?\\[0.5ex]
  7047141	&	Binary?\\[0.5ex]
  7097723	&	Contamination\\
  			&	Cool companion?\\[0.5ex]
  7809801	&	Contamination\\[0.5ex]
  7974841	&	Not flaring\\
  			&	Cool companion?\\[0.5ex]
  7978512	&	-\\[0.5ex]
  8044889	&	Cool companion\\[0.5ex]
  8351193	&	Overlapping neighbour?\\
  			&	Not flaring?\\[0.5ex]
  8367661	&	Instrumental\\[0.5ex]
  8686824	&	Contamination\\[0.5ex]
  8703413	&	Not flaring\\
  			&	Cool companion?\\[0.5ex]
  9216367	&	Cool companion\\
  			&	Overlapping neighbour?\\[0.5ex]
  9782810	&	Overlapping neighbour?\\[0.5ex]
  10082844	&	Binary?\\[0.5ex]
  10489286	&	Overlapping neighbour?\\[0.5ex]
  10817620	&	Binary?\\[0.5ex]
  10971633	&	Cool companion\\
  			&	Overlapping neighbour?\\[0.5ex]
  10974032	&	Cool companion\\[0.5ex]
  11189959	&	Cool companion\\[0.5ex]
  11236035	&	Binary?\\[0.5ex]
  11443271	&	Instrumental\\[0.5ex]
  11600717	& 	Not flaring\\[0.5ex]
  12061741	&	Cool companion\\
  			&	Overlapping neighbour?\\[0.5ex]
	\hline
	\end{tabular}
	\end{center}
\end{table}

Table \ref{Tab:Explanations} gives a summary of our results. Stars for which flare-like features were detected but excluded after comparison with other lightcurves are listed as having an `Instrumental' origin. If stars in either of the two categories are also found to be SB1s, we add `Cool companion?' as another possible explanation. Flares for stars in binary systems are listed as originating from a `Cool companion', which we have shown above can explain the presence of the flares. Stars with unconfirmed binarity features listed in Table \ref{Tab:Fourier} are marked by `Binary?' in Table \ref{Tab:Explanations}. Clear contamination from nearby stars is noted as `Contamination' as explanation for the flaring when valid. Furthermore, we mark stars with `Overlapping neighbours' since the lightcurves of such stars will clearly be influenced by the neighbouring star. 

Through our analysis we have not found an example of an intrinsicly flaring A--type star, for which we can rule out all other possible explanations. If any such star exists, it should be among the nine stars (including unconfirmed binaries) for which we have not yet found or explored all of the possible explanations for the observed flaring. Stars of special interest are single stars and/or stars with H$\alpha$ emission. KIC 7047141 is of particular interest as it has the strongest H$\alpha$ emission in our sample, it is isolated and not contaminated by any nearby source. However, at the moment we do not have sufficient data available to determine whether this star is in a binary system or not.  

\section{Conclusions}\label{Sec:Conclusions}

Possessing only shallow convective envelopes, weak large scale magnetic fields and weak stellar winds, A--type stars should be unable to support any of the known flare mechanisms. In spite of this \citet{Balona2012,Balona2013} has detected flare-like events in the lightcurves of 33 A--type stars. In order to investigate the origin of these flares, we have carried out a stringent flare detection using similar detection criteria as in \citet{Walkowicz2011} with the aim to confirm the presence of these features. In the process, we considered the entire available \emph{Kepler} dataset from Q0-Q17 for the stars. While there are large discrepancies between the number of detected flares in this work and in \citet{Balona2012,Balona2013}, we have confirmed that 27 out of the 33 A--type stars do indeed show such events in their lightcurves.

For the 27 flaring A--type stars, we studied the FOV in three different filters and the pixel data in order to look for possible sources of contamination of the lightcurves. We found that 14 of the stars have overlapping neighbouring stars in their FOV, which will therefore have a clear influence on the lightcurves depending on the differences in apparent magnitudes. Furthermore, five stars show clear pixel contamination from close or overlapping neighbours. 

The analysis of the FIES spectra of 22 of the initial 33 stars has confirmed that all stars lie within the expected activity quiet range from B5-F5. Furthermore, we find none of the characteristic features of Ap/Bp type stars in their spectra.

We determined $RV$s for 22 stars through cross-correlation with computed synthetic spectra. Combining these results with $RV$s available from literature, we have $RV$s for at least two epochs for 19 of the stars. Of these 19 stars we find 11 to be SB1s, five single stars and three unknown. All of the detected SB1s show binarity features in their lightcurves and Fourier spectra. The Fourier analysis revealed that one of the SB1s is in fact a heartbeat star. A comparison of the phase folded lightcurve with the flare timings in this system show no indication that the flares originate from an interaction between the two stars. For the eight stars out of the 11 SB1s that we find to be flaring, we estimated the flare energies and amplitudes for the most energetic flare in each star if these originated from a cooler companion. We find that the flares can in fact stem from a cooler companion, contradicting the conclusions of \citet{Balona2012,Balona2013}.

While analysing the FIES spectra, we noticed the presence of narrow absorption features in the \ion{Ca}{ii} K and \ion{NaD}{} lines of at least 11 stars. Following the procedure in \citet{Vallerga1993} we fitted these features as cloud components and determined their $RV$s. Comparing these to the stellar $RV$s we find that these features are more likely interstellar rather than circumstellar in nature. Eight stars show H$\alpha$ emission in their spectra and one star (KIC 7047141) appears to have multiple Balmer lines in emission. For KIC 7978512 the H$\alpha$ emission varies from one epoch to the next. Whether this observed emission is circumstellar or magnetic in origin is unknown.

To summarize we find possible, non-intrinsic explanations for at least 19 out of the 33 A--type stars by \citet{Balona2012,Balona2013}. Nine stars, six of which do not have any available FIES spectra, require further analysis before any conclusions can be made. Our analyses so far, have not identified any of the target stars to clearly be flaring, casting serious doubts on the flaring A-type stars hypothesis.

\section*{Acknowledgements}

Based on observations made with the Nordic Optical Telescope, operated by the Nordic Optical Telescope Scientific Association at the Observatorio del Roque de los Muchachos, La Palma, Spain, of the Instituto de Astrofisica de Canarias. October 2013 FIES spectra were obtained during the course `Practical observing course with Nordic Optical Telescope'. We thank the Finnish Centre for Astronomy with ESO for organising this course. This paper includes data collected by the \emph{Kepler} mission. Funding for the \emph{Kepler} mission is provided by the NASA Science Mission Directorate.

Funding for the Stellar Astrophysics Centre is provided by The Danish National Research Foundation (Grant DNRF106). The research was supported by the ASTERISK project (ASTERoseismic Investigations with SONG and Kepler) funded by the European Research Council (Grant agreement no.: 267864). Part of the research leading to these results has received funding from the European Research Council (ERC) under the European Union’s Horizon 2020 research and innovation programme (grant agreement N$^\circ$670519: MAMSIE). TRW acknowledges support from the Villum Foundation (research grant 10118).





\bibliographystyle{mnras}
\bibliography{references.bib} 

\begin{thebibliography}{}
\makeatletter
\relax
\def\mn@urlcharsother{\let\do\@makeother \do\$\do\&\do\#\do\^\do\_\do\%\do\~}
\def\mn@doi{\begingroup\mn@urlcharsother \@ifnextchar [ {\mn@doi@}
  {\mn@doi@[]}}
\def\mn@doi@[#1]#2{\def\@tempa{#1}\ifx\@tempa\@empty \href
  {http://dx.doi.org/#2} {doi:#2}\else \href {http://dx.doi.org/#2} {#1}\fi
  \endgroup}
\def\mn@eprint#1#2{\mn@eprint@#1:#2::\@nil}
\def\mn@eprint@arXiv#1{\href {http://arxiv.org/abs/#1} {{\tt arXiv:#1}}}
\def\mn@eprint@dblp#1{\href {http://dblp.uni-trier.de/rec/bibtex/#1.xml}
  {dblp:#1}}
\def\mn@eprint@#1:#2:#3:#4\@nil{\def\@tempa {#1}\def\@tempb {#2}\def\@tempc
  {#3}\ifx \@tempc \@empty \let \@tempc \@tempb \let \@tempb \@tempa \fi \ifx
  \@tempb \@empty \def\@tempb {arXiv}\fi \@ifundefined
  {mn@eprint@\@tempb}{\@tempb:\@tempc}{\expandafter \expandafter \csname
  mn@eprint@\@tempb\endcsname \expandafter{\@tempc}}}

\bibitem[\protect\citeauthoryear{{Allen}}{{Allen}}{1973}]{Allen1973}
{Allen} C.~W.,  1973, {Astrophysical quantities}.
University of London, Athlone Press

\bibitem[\protect\citeauthoryear{{Balona}}{{Balona}}{2012}]{Balona2012}
{Balona} L.~A.,  2012, \mn@doi [\mnras] {10.1111/j.1365-2966.2012.21135.x},
  \href {http://adsabs.harvard.edu/abs/2012MNRAS.423.3420B} {423, 3420}

\bibitem[\protect\citeauthoryear{{Balona}}{{Balona}}{2013}]{Balona2013}
{Balona} L.~A.,  2013, \mn@doi [\mnras] {10.1093/mnras/stt322}, \href
  {http://adsabs.harvard.edu/abs/2013MNRAS.431.2240B} {431, 2240}

\bibitem[\protect\citeauthoryear{{Bertaux}, {Lallement}, {Ferron}  \&
  {Boonne}}{{Bertaux} et~al.}{2014a}]{Bertaux2014b}
{Bertaux} J.-L.,  {Lallement} R.,  {Ferron} S.,   {Boonne} C.,  2014a, in 13th
  International HITRAN Conference. , \mn@doi{10.5281/zenodo.11110}

\bibitem[\protect\citeauthoryear{{Bertaux}, {Lallement}, {Ferron}, {Boonne}  \&
  {Bodichon}}{{Bertaux} et~al.}{2014b}]{Bertaux2014}
{Bertaux} J.~L.,  {Lallement} R.,  {Ferron} S.,  {Boonne} C.,   {Bodichon} R.,
  2014b, \mn@doi [\aap] {10.1051/0004-6361/201322383}, \href
  {http://adsabs.harvard.edu/abs/2014A%26A...564A..46B} {564, A46}

\bibitem[\protect\citeauthoryear{{Boch} \& {Fernique}}{{Boch} \&
  {Fernique}}{2014}]{Boch2014}
{Boch} T.,  {Fernique} P.,  2014, in {Manset} N.,  {Forshay} P.,  eds,
  Astronomical Society of the Pacific Conference Series Vol. 485, Astronomical
  Data Analysis Software and Systems XXIII. p.~277

\bibitem[\protect\citeauthoryear{{Bonnarel} et~al.,}{{Bonnarel}
  et~al.}{2000}]{Bonnarel2000}
{Bonnarel} F.,  et~al., 2000, \mn@doi [\aaps] {10.1051/aas:2000331}, \href
  {http://cdsads.u-strasbg.fr/abs/2000A%26AS..143...33B} {143, 33}

\bibitem[\protect\citeauthoryear{{Braithwaite}}{{Braithwaite}}{2014}]{Braithwaite2014}
{Braithwaite} J.,  2014, in {Petit} P.,  {Jardine} M.,   {Spruit} H.~C.,  eds,
  IAU Symposium Vol. 302, Magnetic Fields throughout Stellar Evolution. pp
  255--264 (\mn@eprint {arXiv} {1312.4755}), \mn@doi{10.1017/S1743921314002221}

\bibitem[\protect\citeauthoryear{{Briquet}, {Korhonen}, {Gonz{\'a}lez},
  {Hubrig}  \& {Hackman}}{{Briquet} et~al.}{2010}]{Briquet2010}
{Briquet} M.,  {Korhonen} H.,  {Gonz{\'a}lez} J.~F.,  {Hubrig} S.,   {Hackman}
  T.,  2010, \mn@doi [\aap] {10.1051/0004-6361/200913775}, \href
  {http://adsabs.harvard.edu/abs/2010A%26A...511A..71B} {511, A71}

\bibitem[\protect\citeauthoryear{{Brown}, {Latham}, {Everett}  \&
  {Esquerdo}}{{Brown} et~al.}{2011}]{Brown2011}
{Brown} T.~M.,  {Latham} D.~W.,  {Everett} M.~E.,   {Esquerdo} G.~A.,  2011,
  \mn@doi [\aj] {10.1088/0004-6256/142/4/112}, \href
  {http://adsabs.harvard.edu/abs/2011AJ....142..112B} {142, 112}

\bibitem[\protect\citeauthoryear{{Bryson} et~al.,}{{Bryson}
  et~al.}{2010}]{Bryson2010a}
{Bryson} S.~T.,  et~al., 2010, in Software and Cyberinfrastructure for
  Astronomy. p. 77401D, \mn@doi{10.1117/12.857625}

\bibitem[\protect\citeauthoryear{{Capelli}, {Warwick}, {Cappelluti},
  {Gillessen}, {Predehl}, {Porquet}  \& {Czesla}}{{Capelli}
  et~al.}{2011}]{Capelli2011}
{Capelli} R.,  {Warwick} R.~S.,  {Cappelluti} N.,  {Gillessen} S.,  {Predehl}
  P.,  {Porquet} D.,   {Czesla} S.,  2011, \mn@doi [\aap]
  {10.1051/0004-6361/201015758}, \href
  {http://adsabs.harvard.edu/abs/2011A%26A...525L...2C} {525, L2}

\bibitem[\protect\citeauthoryear{{Cardelli} \& {Wallerstein}}{{Cardelli} \&
  {Wallerstein}}{1986}]{Cardelli1986}
{Cardelli} J.~A.,  {Wallerstein} G.,  1986, \mn@doi [\apj] {10.1086/164008},
  \href {http://adsabs.harvard.edu/abs/1986ApJ...302..492C} {302, 492}

\bibitem[\protect\citeauthoryear{{Carrington}}{{Carrington}}{1859}]{Carrington1859}
{Carrington} R.~C.,  1859, \mn@doi [\mnras] {10.1093/mnras/20.1.13}, \href
  {http://adsabs.harvard.edu/abs/1859MNRAS..20...13C} {20, 13}

\bibitem[\protect\citeauthoryear{{Castelli} \& {Kurucz}}{{Castelli} \&
  {Kurucz}}{2004}]{Castelli2004}
{Castelli} F.,  {Kurucz} R.~L.,  2004, ArXiv Astrophysics e-prints, \href
  {http://adsabs.harvard.edu/abs/2004astro.ph..5087C} {}

\bibitem[\protect\citeauthoryear{{Catanzaro} \& {Ripepi}}{{Catanzaro} \&
  {Ripepi}}{2014}]{Catanzaro2014}
{Catanzaro} G.,  {Ripepi} V.,  2014, \mn@doi [\mnras] {10.1093/mnras/stu674},
  \href {http://cdsads.u-strasbg.fr/abs/2014MNRAS.441.1669C} {441, 1669}

\bibitem[\protect\citeauthoryear{{Charbonneau}}{{Charbonneau}}{2010}]{Charbonneau2010}
{Charbonneau} P.,  2010, Living Rev. Solar Phys., \href
  {http://www.livingreviews.org/lrsp-2010-3} {7, 3}

\bibitem[\protect\citeauthoryear{{Christiansen} et~al.,}{{Christiansen}
  et~al.}{2013}]{ChristiansenVanCleve2011}
{Christiansen} J.~L.,  et~al., 2013, {Kepler Data Characteristics Handbook,
  KSCI 19040-004}.
NASA Ames Research Center

\bibitem[\protect\citeauthoryear{{Coughlin} et~al.,}{{Coughlin}
  et~al.}{2014}]{Coughlin2014}
{Coughlin} J.~L.,  et~al., 2014, \mn@doi [\aj] {10.1088/0004-6256/147/5/119},
  \href {http://adsabs.harvard.edu/abs/2014AJ....147..119C} {147, 119}

\bibitem[\protect\citeauthoryear{{Dudorov} \& {Khaibrakhmanov}}{{Dudorov} \&
  {Khaibrakhmanov}}{2015}]{Dudorov2015}
{Dudorov} A.~E.,  {Khaibrakhmanov} S.~A.,  2015, \mn@doi [Advances in Space
  Research] {10.1016/j.asr.2014.05.034}, \href
  {http://adsabs.harvard.edu/abs/2015AdSpR..55..843D} {55, 843}

\bibitem[\protect\citeauthoryear{{Duflot}, {Figon}  \& {Meyssonnier}}{{Duflot}
  et~al.}{1995}]{Duflot1995}
{Duflot} M.,  {Figon} P.,   {Meyssonnier} N.,  1995, \aaps, \href
  {http://cdsads.u-strasbg.fr/abs/1995A%26AS..114..269D} {114, 269}

\bibitem[\protect\citeauthoryear{{Fehrenbach} \& {Burnage}}{{Fehrenbach} \&
  {Burnage}}{1990}]{Fehrenbach1990}
{Fehrenbach} C.,  {Burnage} R.,  1990, \aaps, \href
  {http://cdsads.u-strasbg.fr/abs/1990A%26AS...83...91F} {83, 91}

\bibitem[\protect\citeauthoryear{{Fehrenbach}, {Duflot}, {Mannone}, {Burnage}
  \& {Genty}}{{Fehrenbach} et~al.}{1997}]{Fehrenbach1997}
{Fehrenbach} C.,  {Duflot} M.,  {Mannone} C.,  {Burnage} R.,   {Genty} V.,
  1997, \aaps, \href {http://adsabs.harvard.edu/abs/1997A%26AS..124..255F}
  {124, 255}

\bibitem[\protect\citeauthoryear{{Floquet}}{{Floquet}}{1975}]{Floquet1975}
{Floquet} M.,  1975, \aaps, \href
  {http://cdsads.u-strasbg.fr/abs/1975A%26AS...21...25F} {21, 25}

\bibitem[\protect\citeauthoryear{{Gontcharov}}{{Gontcharov}}{2006}]{Gontcharov2006}
{Gontcharov} G.~A.,  2006, \mn@doi [Astronomy Letters]
  {10.1134/S1063773706110065}, \href
  {http://cdsads.u-strasbg.fr/abs/2006AstL...32..759G} {32, 759}

\bibitem[\protect\citeauthoryear{{Gray} \& {Corbally}}{{Gray} \&
  {Corbally}}{1994}]{Gray1994}
{Gray} R.~O.,  {Corbally} C.~J.,  1994, \mn@doi [\aj] {10.1086/116893}, \href
  {http://adsabs.harvard.edu/abs/1994AJ....107..742G} {107, 742}

\bibitem[\protect\citeauthoryear{{Gray} \& {Corbally}}{{Gray} \&
  {Corbally}}{2009}]{GrayCorbally}
{Gray} R.~O.,  {Corbally} C.~J.,  2009, {Stellar Spectral Classification}.
Princeton University Press

\bibitem[\protect\citeauthoryear{{Grenier} et~al.,}{{Grenier}
  et~al.}{1999}]{Grenier1999}
{Grenier} S.,  et~al., 1999, \mn@doi [\aaps] {10.1051/aas:1999489}, \href
  {http://cdsads.u-strasbg.fr/abs/1999A%26AS..137..451G} {137, 451}

\bibitem[\protect\citeauthoryear{{Groote} \& {Schmitt}}{{Groote} \&
  {Schmitt}}{2004}]{Groote2004}
{Groote} D.,  {Schmitt} J.~H.~M.~M.,  2004, \mn@doi [\aap]
  {10.1051/0004-6361:20034300}, \href
  {http://adsabs.harvard.edu/abs/2004A%26A...418..235G} {418, 235}

\bibitem[\protect\citeauthoryear{{Hamaguchi}, {Terada}, {Bamba}  \&
  {Koyama}}{{Hamaguchi} et~al.}{2000}]{Hamaguchi2000}
{Hamaguchi} K.,  {Terada} H.,  {Bamba} A.,   {Koyama} K.,  2000, \mn@doi [\apj]
  {10.1086/308607}, \href {http://adsabs.harvard.edu/abs/2000ApJ...532.1111H}
  {532, 1111}

\bibitem[\protect\citeauthoryear{{Hobbs}, {Vidal-Madjar}, {Ferlet}, {Albert}
  \& {Gry}}{{Hobbs} et~al.}{1985}]{Hobbs1985}
{Hobbs} L.~M.,  {Vidal-Madjar} A.,  {Ferlet} R.,  {Albert} C.~E.,   {Gry} C.,
  1985, \mn@doi [\apjl] {10.1086/184485}, \href
  {http://adsabs.harvard.edu/abs/1985ApJ...293L..29H} {293, L29}

\bibitem[\protect\citeauthoryear{{Hodgson}}{{Hodgson}}{1859}]{Hodgson1859}
{Hodgson} R.,  1859, \mn@doi [\mnras] {10.1093/mnras/20.1.15}, \href
  {http://adsabs.harvard.edu/abs/1859MNRAS..20...15H} {20, 15}

\bibitem[\protect\citeauthoryear{{Koch} et~al.,}{{Koch}
  et~al.}{2010}]{Koch2010}
{Koch} D.~G.,  et~al., 2010, \mn@doi [\apjl] {10.1088/2041-8205/713/2/L79},
  \href {http://adsabs.harvard.edu/abs/2010ApJ...713L..79K} {713, L79}

\bibitem[\protect\citeauthoryear{{Kochukhov}, {Adelman}, {Gulliver}  \&
  {Piskunov}}{{Kochukhov} et~al.}{2007}]{Kochukhov2007}
{Kochukhov} O.,  {Adelman} S.~J.,  {Gulliver} A.~F.,   {Piskunov} N.,  2007,
  \mn@doi [Nature Physics] {10.1038/nphys648}, \href
  {http://adsabs.harvard.edu/abs/2007NatPh...3..526K} {3, 526}

\bibitem[\protect\citeauthoryear{{Lenz} \& {Breger}}{{Lenz} \&
  {Breger}}{2005}]{Lenz2005}
{Lenz} P.,  {Breger} M.,  2005, {A Magneto-Kinematic Model of the Solar Cycle}

\bibitem[\protect\citeauthoryear{{Ligni{\`e}res}, {Petit}, {B{\"o}hm}  \&
  {Auri{\`e}re}}{{Ligni{\`e}res} et~al.}{2009}]{Lignieres2009}
{Ligni{\`e}res} F.,  {Petit} P.,  {B{\"o}hm} T.,   {Auri{\`e}re} M.,  2009,
  \mn@doi [\aap] {10.1051/0004-6361/200911996}, \href
  {http://adsabs.harvard.edu/abs/2009A%26A...500L..41L} {500, L41}

\bibitem[\protect\citeauthoryear{{Mamajek}}{{Mamajek}}{2012}]{Mamajek2012}
{Mamajek} E.~E.,  2012, \mn@doi [\apjl] {10.1088/2041-8205/754/2/L20}, \href
  {http://adsabs.harvard.edu/abs/2012ApJ...754L..20M} {754, L20}

\bibitem[\protect\citeauthoryear{{McDonald}, {Zijlstra}  \& {Boyer}}{{McDonald}
  et~al.}{2012}]{McDonald2012}
{McDonald} I.,  {Zijlstra} A.~A.,   {Boyer} M.~L.,  2012, \mn@doi [\mnras]
  {10.1111/j.1365-2966.2012.21873.x}, \href
  {http://cdsads.u-strasbg.fr/abs/2012MNRAS.427..343M} {427, 343}

\bibitem[\protect\citeauthoryear{{Moffat} \& {Corcoran}}{{Moffat} \&
  {Corcoran}}{2009}]{Moffat2009}
{Moffat} A.~F.~J.,  {Corcoran} M.~F.,  2009, \mn@doi [\apj]
  {10.1088/0004-637X/707/1/693}, \href
  {http://adsabs.harvard.edu/abs/2009ApJ...707..693M} {707, 693}

\bibitem[\protect\citeauthoryear{{Neiner} \& {Lampens}}{{Neiner} \&
  {Lampens}}{2015}]{Neiner2015}
{Neiner} C.,  {Lampens} P.,  2015, \mn@doi [\mnras] {10.1093/mnrasl/slv130},
  \href {http://adsabs.harvard.edu/abs/2015MNRAS.454L..86N} {454, L86}

\bibitem[\protect\citeauthoryear{{Owocki}, {Townsend}  \& {Ud-Doula}}{{Owocki}
  et~al.}{2007}]{Owocki2007}
{Owocki} S.,  {Townsend} R.,   {Ud-Doula} A.,  2007, \mn@doi [Physics of
  Plasmas] {10.1063/1.2472340}, \href
  {http://adsabs.harvard.edu/abs/2007PhPl...14e6502O} {14, 056502}

\bibitem[\protect\citeauthoryear{{Petit} et~al.,}{{Petit}
  et~al.}{2011}]{Petit2011}
{Petit} P.,  et~al., 2011, \mn@doi [\aap] {10.1051/0004-6361/201117573}, \href
  {http://adsabs.harvard.edu/abs/2011A%26A...532L..13P} {532, L13}

\bibitem[\protect\citeauthoryear{{Pittard} \& {Corcoran}}{{Pittard} \&
  {Corcoran}}{2002}]{Pittard2002}
{Pittard} J.~M.,  {Corcoran} M.~F.,  2002, \mn@doi [\aap]
  {10.1051/0004-6361:20020025}, \href
  {http://adsabs.harvard.edu/abs/2002A%26A...383..636P} {383, 636}

\bibitem[\protect\citeauthoryear{{Power}, {Wade}, {Hanes}, {Aurier}  \&
  {Silvester}}{{Power} et~al.}{2007}]{Power2007}
{Power} J.,  {Wade} G.~A.,  {Hanes} D.~A.,  {Aurier} M.,   {Silvester} J.,
  2007, in {Romanyuk} I.~I.,  {Kudryavtsev} D.~O.,  {Neizvestnaya} O.~M.,
  {Shapoval} V.~M.,  eds, Physics of Magnetic Stars. pp 89--97 (\mn@eprint {}
  {astro-ph/0612557})

\bibitem[\protect\citeauthoryear{{Robrade} \& {Schmitt}}{{Robrade} \&
  {Schmitt}}{2011}]{Robrade2011}
{Robrade} J.,  {Schmitt} J.~H.~M.~M.,  2011, \mn@doi [\aap]
  {10.1051/0004-6361/201116843}, \href
  {http://adsabs.harvard.edu/abs/2011A%26A...531A..58R} {531, A58}

\bibitem[\protect\citeauthoryear{{Rosner} \& {Vaiana}}{{Rosner} \&
  {Vaiana}}{1980}]{Rosner1980}
{Rosner} R.,  {Vaiana} G.~S.,  1980, in {Giacconi} R.,  {Setti} G.,  eds,  NATO
  Advanced Science Institutes (ASI) Series C Vol. 60, NATO Advanced Science
  Institutes (ASI) Series C. pp 129--151

\bibitem[\protect\citeauthoryear{{Sanz-Forcada}, {Franciosini}  \&
  {Pallavicini}}{{Sanz-Forcada} et~al.}{2004}]{Sanz-Forcada2004}
{Sanz-Forcada} J.,  {Franciosini} E.,   {Pallavicini} R.,  2004, \mn@doi [\aap]
  {10.1051/0004-6361:20047159}, \href
  {http://adsabs.harvard.edu/abs/2004A%26A...421..715S} {421, 715}

\bibitem[\protect\citeauthoryear{{Shibata} \& {Magara}}{{Shibata} \&
  {Magara}}{2011}]{Shibata2011}
{Shibata} K.,  {Magara} T.,  2011, Living Rev. Solar Phys., \href
  {http://www.livingreviews.org/lrsp-2011-6} {8, 6}

\bibitem[\protect\citeauthoryear{{Shibayama} et~al.,}{{Shibayama}
  et~al.}{2013}]{Shibayama2013}
{Shibayama} T.,  et~al., 2013, \mn@doi [\apjs] {10.1088/0067-0049/209/1/5},
  \href {http://adsabs.harvard.edu/abs/2013ApJS..209....5S} {209, 5}

\bibitem[\protect\citeauthoryear{{Skrutskie} et~al.,}{{Skrutskie}
  et~al.}{2006}]{Skrutskie2006}
{Skrutskie} M.~F.,  et~al., 2006, \mn@doi [\aj] {10.1086/498708}, \href
  {http://adsabs.harvard.edu/abs/2006AJ....131.1163S} {131, 1163}

\bibitem[\protect\citeauthoryear{{Slettebak}}{{Slettebak}}{1975}]{Slettebak1975}
{Slettebak} A.,  1975, \mn@doi [\apj] {10.1086/153493}, \href
  {http://adsabs.harvard.edu/abs/1975ApJ...197..137S} {197, 137}

\bibitem[\protect\citeauthoryear{{Slettebak}}{{Slettebak}}{1982}]{Slettebak1982}
{Slettebak} A.,  1982, \mn@doi [\apjs] {10.1086/190820}, \href
  {http://adsabs.harvard.edu/abs/1982ApJS...50...55S} {50, 55}

\bibitem[\protect\citeauthoryear{{Smith}, {Grady}, {Peters}  \&
  {Feigelson}}{{Smith} et~al.}{1993}]{Smith1993}
{Smith} M.~A.,  {Grady} C.~A.,  {Peters} G.~J.,   {Feigelson} E.~D.,  1993,
  \mn@doi [\apjl] {10.1086/186857}, \href
  {http://adsabs.harvard.edu/abs/1993ApJ...409L..49S} {409, L49}

\bibitem[\protect\citeauthoryear{{Steck}}{{Steck}}{2008}]{Steck2008}
{Steck} D.~A.,  2008, {Sodium D Line Data}, available online at
  \url{http://steck.us/alkalidata}

\bibitem[\protect\citeauthoryear{{Stelzer}, {Flaccomio}, {Montmerle}, {Micela},
  {Sciortino}, {Favata}, {Preibisch}  \& {Feigelson}}{{Stelzer}
  et~al.}{2005}]{Stelzer2005}
{Stelzer} B.,  {Flaccomio} E.,  {Montmerle} T.,  {Micela} G.,  {Sciortino} S.,
  {Favata} F.,  {Preibisch} T.,   {Feigelson} E.~D.,  2005, \mn@doi [\apjs]
  {10.1086/432375}, \href {http://adsabs.harvard.edu/abs/2005ApJS..160..557S}
  {160, 557}

\bibitem[\protect\citeauthoryear{{Telting} et~al.,}{{Telting}
  et~al.}{2014}]{Telting2014}
{Telting} J.~H.,  et~al., 2014, \mn@doi [Astronomische Nachrichten]
  {10.1002/asna.201312007}, \href
  {http://adsabs.harvard.edu/abs/2014AN....335...41T} {335, 41}

\bibitem[\protect\citeauthoryear{{Thompson} et~al.,}{{Thompson}
  et~al.}{2012}]{Thompson2012}
{Thompson} S.~E.,  et~al., 2012, \mn@doi [\apj] {10.1088/0004-637X/753/1/86},
  \href {http://adsabs.harvard.edu/abs/2012ApJ...753...86T} {753, 86}

\bibitem[\protect\citeauthoryear{{Tkachenko}, {Lehmann}, {Smalley}  \&
  {Uytterhoeven}}{{Tkachenko} et~al.}{2013}]{Tkachenko2013}
{Tkachenko} A.,  {Lehmann} H.,  {Smalley} B.,   {Uytterhoeven} K.,  2013,
  \mn@doi [\mnras] {10.1093/mnras/stt453}, \href
  {http://cdsads.u-strasbg.fr/abs/2013MNRAS.431.3685T} {431, 3685}

\bibitem[\protect\citeauthoryear{{Tody}}{{Tody}}{1986}]{Tody1986}
{Tody} D.,  1986, in {Crawford} D.~L.,  ed.,  \procspie Vol. 627,
  Instrumentation in astronomy VI. p.~733

\bibitem[\protect\citeauthoryear{{Tody}}{{Tody}}{1993}]{Tody1993}
{Tody} D.,  1993, in {Hanisch} R.~J.,  {Brissenden} R.~J.~V.,   {Barnes} J.,
  eds,  Astronomical Society of the Pacific Conference Series Vol. 52,
  Astronomical Data Analysis Software and Systems II. p.~173

\bibitem[\protect\citeauthoryear{{Townsend} \& {Owocki}}{{Townsend} \&
  {Owocki}}{2005}]{Townsend2005}
{Townsend} R.~H.~D.,  {Owocki} S.~P.,  2005, \mn@doi [\mnras]
  {10.1111/j.1365-2966.2005.08642.x}, \href
  {http://adsabs.harvard.edu/abs/2005MNRAS.357..251T} {357, 251}

\bibitem[\protect\citeauthoryear{{Vallerga}, {Vedder}, {Craig}  \&
  {Welsh}}{{Vallerga} et~al.}{1993}]{Vallerga1993}
{Vallerga} J.~V.,  {Vedder} P.~W.,  {Craig} N.,   {Welsh} B.~Y.,  1993, \mn@doi
  [\apj] {10.1086/172875}, \href
  {http://adsabs.harvard.edu/abs/1993ApJ...411..729V} {411, 729}

\bibitem[\protect\citeauthoryear{{Van Cleve}, {Caldwell}, {Thompson}, {Haas},
  {Koch}  \& {Borucki}}{{Van Cleve} et~al.}{2009}]{VanCleve2009}
{Van Cleve} J.,  {Caldwell} D.,  {Thompson} R.,  {Haas} M.~R.,  {Koch} D.,
  {Borucki} W.~J.,  2009, {Kepler Instrument Handbook, KSCI 19033-001}.
NASA Ames Research Center

\bibitem[\protect\citeauthoryear{{Verner}, {Bruhweiler}  \& {Gull}}{{Verner}
  et~al.}{2005}]{Verner2005}
{Verner} E.,  {Bruhweiler} F.,   {Gull} T.,  2005, \mn@doi [\apj]
  {10.1086/429400}, \href {http://adsabs.harvard.edu/abs/2005ApJ...624..973V}
  {624, 973}

\bibitem[\protect\citeauthoryear{{Vink}, {O'Neill}, {Els}  \& {Drew}}{{Vink}
  et~al.}{2005}]{Vink2005}
{Vink} J.~S.,  {O'Neill} P.~M.,  {Els} S.~G.,   {Drew} J.~E.,  2005, \mn@doi
  [\aap] {10.1051/0004-6361:200500141}, \href
  {http://cdsads.u-strasbg.fr/abs/2005A%26A...438L..21V} {438, L21}

\bibitem[\protect\citeauthoryear{{Walkowicz} et~al.,}{{Walkowicz}
  et~al.}{2011}]{Walkowicz2011}
{Walkowicz} L.~M.,  et~al., 2011, \mn@doi [\aj] {10.1088/0004-6256/141/2/50},
  \href {http://adsabs.harvard.edu/abs/2011AJ....141...50W} {141, 50}

\bibitem[\protect\citeauthoryear{{Welsh}, {Lallement}, {Vergely}  \&
  {Raimond}}{{Welsh} et~al.}{2010}]{Welsh2010}
{Welsh} B.~Y.,  {Lallement} R.,  {Vergely} J.-L.,   {Raimond} S.,  2010,
  \mn@doi [\aap] {10.1051/0004-6361/200913202}, \href
  {http://adsabs.harvard.edu/abs/2010A%26A...510A..54W} {510, A54}

\bibitem[\protect\citeauthoryear{{Welsh} et~al.,}{{Welsh}
  et~al.}{2011}]{Welsh2011}
{Welsh} W.~F.,  et~al., 2011, \mn@doi [\apjs] {10.1088/0067-0049/197/1/4},
  \href {http://adsabs.harvard.edu/abs/2011ApJS..197....4W} {197, 4}

\makeatother
\end{thebibliography}





\bsp	
\label{lastpage}
\end{document}